\documentclass[12pt]{article}
\usepackage{arxiv}
\usepackage[utf8]{inputenc} 
\usepackage{hyperref}       
\usepackage{url}           
\usepackage[sort]{natbib}
\usepackage{doi}

\usepackage{adjustbox}
\usepackage{tabularx}

\usepackage{booktabs}
 
\renewcommand{\headeright}{}
\usepackage{subfigure}
\usepackage{amsmath}
\usepackage[dvipsnames]{xcolor}
\usepackage{color, soul}
\usepackage{comment}
\usepackage{graphicx}
\usepackage[T1]{fontenc}

\usepackage{longtable}
\usepackage{seqsplit}
\renewcommand{\headeright}{}
\renewcommand{\undertitle}{This preprint has not been peer-reviewed.}
\renewcommand{\shorttitle}{}

\hypersetup{
    bookmarksopen=false,
    unicode=false,          
    pdftoolbar=true,        
    pdfmenubar=true,        
    pdffitwindow=false,     
    pdfstartview={FitH},    
    pdftitle={Inferring Political Leaning from Non-political Content}, 
    pdfauthor={Ahmet Kurnaz, Scott A.\ Hale},     
    pdfnewwindow=true,      
    colorlinks=true,       
    linkcolor=black,          
    citecolor=NavyBlue,        
    filecolor=NavyBlue,      
    urlcolor=NavyBlue           
}

\begin{document}

\title{Top Gear or Black Mirror: Inferring Political Leaning from Non-political Content}
\author{
	\href{https://orcid.org/0000-0001-5628-328X}{\includegraphics[scale=0.06]{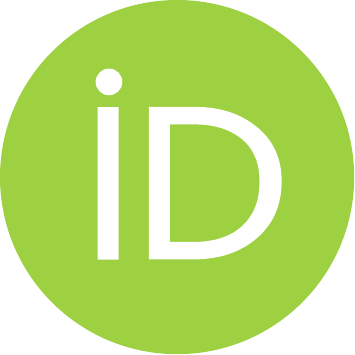}\hspace{1mm}Ahmet ~Kurnaz} \\
	Department of Public Administration and Political Science\\
	Çanakkale Onsekiz Mart University\\
	Çanakkale, Turkey \\
	\texttt{ahmetkurnaz@hotmail.com} \\
	\And
	 \href{https://orcid.org/0000-0002-6894-4951}{\includegraphics[scale=0.06]{orcid.pdf}\hspace{1mm}Scott A. ~Hale}\\
	Oxford Internet Institute\\
	University of Oxford\\
	Oxford, United Kingdom \\
	\texttt{scott.hale@oii.ox.ac.uk} \\
}

\renewcommand{\cite}{\citep}

\maketitle
\begin{abstract}

Polarization and echo chambers are often studied in the context of explicitly political events such as elections, and little scholarship has examined the mixing of political groups in non-political contexts. A major obstacle to studying political polarization in non-political contexts is that political leaning (i.e., left vs\ right orientation) is often unknown.
Nonetheless, political leaning is known to correlate (sometimes quite strongly) with many lifestyle choices leading to stereotypes such as the ``latte-drinking liberal.'' We develop a machine learning classifier to infer political leaning from non-political text and, optionally, the accounts a user follows on social media. We use Voter Advice Application results shared on Twitter as our groundtruth and train and test our classifier on a Twitter dataset comprising the 3,200 most recent tweets of each user after removing any tweets with political text. We correctly classify the political leaning of most users (F1 scores range from 0.70 to 0.85 depending on coverage). We find no relationship between the level of political activity and our classification results. We apply our classifier to a case study of news sharing in the UK and discover that, in general, the sharing of political news exhibits a distinctive left--right divide while sports news does not. 
\end{abstract}

\maketitle

\section{Introduction}

Word choice, grammatical structures, and other linguistic features often correlate with demographic variables such as age and gender, although audience and other performative aspects can play a significant role~\cite{wang_demographic_2019,nguyen2016computational}. 

Latent attribute inference describes the field of research in computer science that tries to infer demographic and other attributes from individuals' behaviours.
While much attention has been paid to inferring gender, age, and location from social media data~\cite{wang_demographic_2019,liu_inferring_2019,zagheni_leveraging_2017,zhang_your_2016,chen_comparative_2015, sap_developing_2014,nguyen_how_2013,rosenthal_age_2011,rao_classifying_2010}, less work has been done on inferring political orientation. 
Nonetheless, political orientation is known to correlate (sometimes quite strongly) with many lifestyle choices, which leads to stereotypes such as the ``latte-drinking liberal'' or ``bird-hunting conservative''~\cite{della2015why}. 
Individuals' perception of their own political 
identities also correlates with the degree of moral value adoption and behaviour \cite{talaifar_deep_2019}. 
Furthermore, political leaning may influence perceptions of non-political topics: \citet{ahn_nonpolitical_2014} were able to infer political leaning from fMRI data in which subjects were shown non-political images as stimuli. Even academic publications and their findings contain some signals of the political leanings of their authors~\cite{jelveh_detecting_2014}. 

It, therefore, is very likely that the content people share, the accounts they follow, and the words they use on social media may also contain traces of their political orientation and that a properly trained machine learning system will be able to infer individuals' political orientations even in non-political contexts.

Although there are several political spectrum models \cite{eysenck_sense_1964, sznajd-weron_who_2005, kitschelt_transformation_1994}, in this study, we focus specifically on a unidimensional left--right spectrum, given its ubiquity in political science and popular culture alike.

We develop and freely share our machine learning system, which achieves an F1 score of more than 0.85 
even when considering non-political content. 
Our ground-truth data come from individuals who used Voter Advice Applications (VAAs) during the 2015 and 2017 UK General Elections. 
Many respondents shared their VAA results on social media during the elections, and we use these social media accounts and their VAA results to train our system. We gather multiple datasets, including these users' political and non-political Twitter activity.

After developing our classifier, we use the system to infer the political orientations of individuals sharing news articles to investigate the level of polarization in  sharing different types of news and from different sources. Specifically, we investigate the sharing of political and sports news from \textit{The Telegraph} (right-leaning), the \textit{BBC} (centrist), and \textit{The Guardian} (left-leaning).

We first present our classifier's motivation, development, and evaluation before turning to the news sharing case study. Finally, we conclude on broader implications and future research directions. 

\section{Inferring political leaning}\label{sec:classifier}

Political ideology prediction from digital trace data has become a core interest for researchers. Prior work has focused on predicting ideological stances using text data, network data, or a combination.

Social media, news outlets, and parliamentary discussions are the most popular sources of textual data: work on social media has used both users' and politicians' textual data \cite{preotiuc-pietro_beyond_2017,conover_predicting_2011}. Newsgroups are also often political, and the political leanings of their text have been studied \cite{kulkarni_multi-view_2018,iyyer_political_2014}. Although legislators are already officially affiliated with a political party, several studies have classified the members of the US Congress~\cite{iyyer_political_2014,diermeier_language_2012,yu_classifying_2008}. 

Text-based approaches focus on the words people choose to use \cite{preotiuc-pietro_beyond_2017}, the style of their messages, and social-media specific attributes such as hashtags \cite{Weber2013PoliticalHT,conover_predicting_2011}. These approaches work well for polarised topics where the left and right use different phrases to talk about the same things: for example, ``gun control'' and ``gun rights.''

Network approaches have used both textual elements such as hashtags, mentions, and URLs \cite{gu_ideology_2016} to create networks, and social links \cite{barbera_birds_2015}.

Within Twitter, there is a divide between using either retweets and mentions \cite[e.g.,][]{conover_predicting_2011,gu_ideology_2016,hale_global_2014} or followers and friends \cite[e.g.,][]{king_ideological_2016,barbera_birds_2015,golbeck_method_2014,pennacchiotti_democrats_2011,zamal_homophily_2012,compton_geotagging_2014} to infer latent attributes. 
Networks based on mentions and retweets are often preferred over followers for two reasons. First, taking a snapshot of an extensive network is difficult, given the Twitter API rate limits. Second, retweeting and mentions, often have greater recency, can be analysed temporally, and can form weighted networks (e.g., based on how many times users mention each other). As retweeting is more publicly visible than following a user, there is also a greater social cost to retweeting another account, suggesting that users will more carefully consider which users they retweet or mention.  
Retweeting has been used as a sign of support \cite[e.g.,][]{conover_predicting_2011,wong_quantifying_2016}, but it is important to note that some mentions, retweets, and hashtag uses will be specifically to call attention to content with which a user disagrees.

As detailed below, we examine the use both words (through topic modeling) in political and non-political text, and network features (the accounts a user follows) to estimate the political leaning of social media users.

We find network features and political text equally reveal political leaning, but network features have lower coverage. Inferring political leaning with non-political text is less accurate, but network data helps in that task and performance is on par with previous work inferring political leaning from other sources.

\subsubsection{Ground truth data}

We collected all mentions of social media accounts and URLs associated with the three most-used voter advice applications (VAAs) during the 2015 and 2017 General Elections in the United Kingdom.\footnote{We used commercial, firehose access for the 2015 election and elevated direct access during the 2017 election. We believe our dataset represents the entire population as closely as possible.}
VAAs ask individuals a series of political and policy issue questions to help them understand their preferences and how these align with political candidates running for office.
Users of VAAs have the opportunity (but not requirement) to share their results on social media.

The three VAAs from which we collected data were ``I Side With You'' (ISW), ``Who Should You Vote For'' (WSYVF), and ``Vote Match'' (VM). We found 4,118 unique users who shared their VAA results: 2,975 of these users came from ISW, 794 from VM, and 387 from WSYVF. Our data only includes original tweets and not retweets of VAA result URLs. Thirty eight users shared results from multiple VAAs.

After collecting the data during the elections, we extracted all VAA result URLs in July 2018, downloaded the content of VAA result pages, and extracted users' matches with political parties. We furthermore obtained the most recent 3,200 public tweets as of January 2019 of each user who shared a VAA result by using the \textit{user\_timeline} endpoint of Twitter's public RESTful API.

All VAAs provided the voter's alignment with individual political parties. WSYVF also gave a left--right score, but the exact details of this score are unknown. Therefore, we develop our own left--right or political leaning score using only the matches to parties, which can be applied consistently across all three VAAs. 

We calculate political leaning scores by subtracting
the user's match to Labour and the Conservatives (the UK's two largest parties).\footnote{We experimented with summing a user's match with left-leaning parties (Labour, SNP, Greens, Plaid-Cymru, Sinn Fein) and subtracting this from the sum of the user's match to right-leaning parties (Conservatives, UKIP, British National). This procedure produced similar results but was less transparent as not all parties are available in all constituencies.} 
We normalize all scores into the [-1,1] interval for each platform before combining data from all the VAAs together.

Users with a political leaning score greater than zero are labelled as ``right-leaning'' while those with scores less than zero are labelled as ``left-leaning.''  We dropped all users with political leaning scores equal to zero.
A small number of users shared results from multiple VAAs. We checked for consistency and removed one inconsistent user from our dataset. We created one political leaning score for users with multiple but consistent political leanings by taking the mean of their political leaning scores from the multiple VAAs.

In total, our ground-truth dataset contains 2,694 unique users from all three VAAs. We collected last 3,200 tweets from 1,921 of these users in January 2019.

\begin{figure}
\centering
\includegraphics{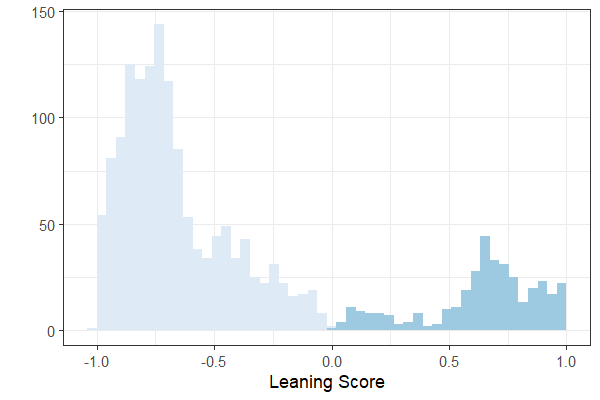}    
\caption{Histogram of left- and right-leaning users sharing Voter Advice Application (VAA) results on Twitter.}
\label{fig:leaning_dist}
\end{figure}

We detect the language of all tweets using the Compact Language Detector 2 \cite{cld2} and remove all users who had less than 75\% of their most recently 3,200 tweets in English. Last, we remove users who tweeted less than 100 times. As a result, the final dataset comprises 1,760 users. Of these users, 1,396 are left-leaning, 
and 364 are right-leaning (Figure~\ref{fig:leaning_dist}).

\subsection{Predictive variables}
We build five datasets using users' tweets and Twitter `friends' (i.e., the accounts a user follows). The first three datasets are (1) network information, (2) political text, and (3) non-political text. The final two-hybrid datasets are constructed by joining textual and network data.

\subsubsection{Textual data}
We create two datasets of texts from tweets authored by the users. The first is a set of \textit{political text} from tweets containing a political term (e.g., \#ge2015, labour) primarily collected during the 2015 and 2017 elections with firehose or elevated streaming API access. The second dataset, \textit{non-political text}, is mainly from users' most recent 3,200 tweets collected with the RESTful API. We find that some recent tweets are political (e.g., discussing Brexit) and exclude tweets with political words from the non-political dataset. 

\subsubsection{Determining political tweets}
To detect political tweets, we develop a two-step procedure. In the first step, we create an index by using election and non-election word occurrence frequencies. Then, we extract initial political words by using this index. Finally, we train a word embedding model to extend the political terms list in the second step. 

We select words that appear in at least 250 different tweets during an election period in the initial step.\footnote{The periods for the elections as follow: GE2010 (2010-04-12 to 2010-06-06), GE2015 (2014-11-19 to 2015-06-07), and GE2017 (2017-04-18 to 2017-07-08).} Then, we examine word occurrences both within and outside of the election periods. We normalize these two frequencies by using the total tweet count for each period. Finally, we divide the out-of-election number by the in-election number to develop a political index.   

In our case, setting political index threshold of 0.25 yields a word list comprising 316 words with scores lower than 0.25. However, this initial list contains several ambiguous words. Therefore, we manually examined all words and removed 78 vague terms (\#eurovision2015, \#pensioners, youtuber, tuition, etc.) before continuing to the second step. We also add political words that were present throughout the whole  period, including `brexit', `eu', `trump', `clinton', `sanders', `johnson', `gop', and `referendum.'

In the second step, we use word embeddings to determine the closest terms to the political words extracted in the first step. We train a skip-gram embedding model (skip window: 5, minimum frequency: 100). 
Using Euclidean distance, we choose the closest three words to any political term in the list. Then, we again manually inspect all words and remove ambiguous terms. This results in 433 political 
words. The complete list of words is available within the supplemental materials.

\begin{figure}
    \centering
    \includegraphics{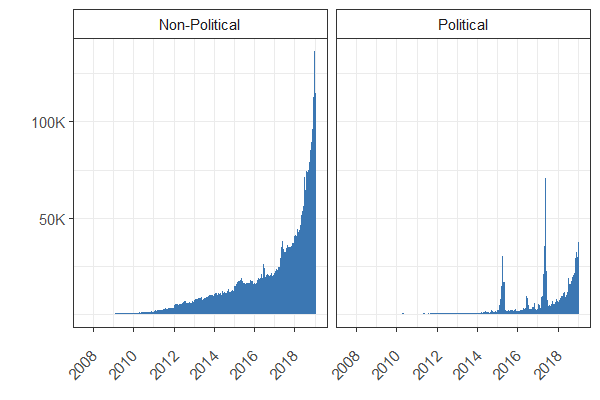}
    \caption{Frequency of political and non-political tweets by date. There are distinctive spikes during the elections for political tweets, but also political tweets outside of this.}
    \label{fig:tweet-timing}
\end{figure}

We label any tweet containing a term from this final list as political. Other tweets are labelled non-political. 
In the end, the non-political dataset consists of 3.66 million tweets, and the political dataset contains 888 thousand tweets (Figure \ref{fig:tweet-timing}).

\subsubsection{Network data}
We collected the accounts that users in our dataset followed (their `friends') on Twitter for 1,742 of the 1,760 users in our dataset who shared VAA results. Next, we create a binary matrix where the columns are Twitter accounts followed by multiple users, and the rows are the users in our dataset.

\subsubsection{Approach}
Our goal is to estimate political leaning at a per-user level, not at a per-tweet level.
We, therefore, sort all tweets of each user by date and concatenate these tweets into two text documents: one for political text and one for non-political text. 

To construct document-feature matrices (DFM), we remove URLs and non-alpha characters from the text.
Next, we stem all words using Porter's algorithm \cite{porter_algorithm_1980} and exclude words from the SMART stop word list \cite{lewis_rcv1:_2004}.
Finally, we create lists of unigram, bigram and trigrams as input features to our classifier. Consequently, our features consist of unigrams, bigrams and trigrams altogether, and the documents are combined tweets of each user.

To reduce the dimension of features, we remove sparse terms 
by setting sparsity for \textit{political dataset} to 0.9 and for \textit{non-political dataset} to 0.85. Consequently, the political DFM consists of 1,760 users and 4,676 features, and the non-political DFM contains 1,760 users and 6,613 features. We also remove sparse terms from the network DFM by setting sparsity to 0.88. Our network DFM consists of 1,742 users and 55 features (followed accounts). 

Last but not least, our dataset contains more left-leaning than right-leaning users. Therefore we select all of the right-leaning users and an equal number of left-leaning users randomly. We combine these two subsets and shuffle the documents in the resulting DFM, which contains 728 users balanced across our two classes (left-leaning and right-leaning) in the political and non-political sample datasets. When using network data as an input on its own or in combination with text, the numbers are slightly lower as network data was not available for all users.

\begin{table}[tb]
\centering
        \begin{tabular}{lccccc}
        \toprule
            Dataset 
            &NB
            &NN
            &SVM\textsubscript{lin}
            &SVM\textsubscript{poly}
            &SVM\textsubscript{rad}
            \\
            \midrule
            net 
             & 0.27 & 0.75 & 0.73 & 0.75 & 0.75    \\
       
            non-pol
             &0.54 & 0.64 & 0.60 & 0.65 & 0.65    \\
          
            non-pol+net
             &0.59 & 0.70 & 0.68 & 0.71 & 0.72  \\
            
            pol 
             &0.66 & 0.75 & 0.74 & 0.73 & 0.72  \\
          
            pol+net
             & 0.67 & 0.69 & 0.66 & 0.71 & 0.67 \\
            \bottomrule\\
        \end{tabular}
        \caption{Average F1 scores for the training datasets. We use 
        10-fold cross-validation while training the models. We repeat this procedure for five different randomly selected, but balanced, samples.}
        \label{tbl:auc_results}
\end{table}

\begin{table*}[tb]
        \begin{adjustbox}{width=\textwidth}
        \begin{tabular}{l ccc ccc ccc ccc ccc}
        \toprule
             
            & \multicolumn{3}{c}{network}
            & \multicolumn{3}{c}{non-political}
            & \multicolumn{3}{c}{non-pol+net}
            & \multicolumn{3}{c}{political}
            & \multicolumn{3}{c}{pol+net}
            \\
            \cmidrule(lr){2-4}
            \cmidrule(lr){5-7}
            \cmidrule(lr){8-10}
            \cmidrule(lr){11-13}
            \cmidrule(l){14-16}
            
            Classifier
            & F1 & P & R
            & F1 & P & R
            & F1 & P & R
            & F1 & P & R
            & F1 & P & R\\
            \midrule

Na\"ive Bayes & 0.29 & 0.17 & \textbf{1.00} & 0.64 & 0.63 & 0.67 & 0.70 & 0.69 & 0.72 & 0.75 & 0.79 & 0.71 & 0.73 & \textbf{0.75} & 0.71\\

Neural Network & 0.77 & 0.71 & 0.83 & 0.69 & 0.67 & 0.73 & 0.77 & \textbf{0.74} & 0.80 & \textbf{0.81} & \textbf{0.80} & 0.82 & 0.74 & 0.70 & 0.78\\

SVM\textsubscript{lin} & 0.74 & 0.66 & 0.85 & 0.65 & 0.59 & \textbf{0.78} & 0.72 & 0.66 & 0.80 & 0.79 & 0.76 & \textbf{0.83} & 0.74 & 0.68 & 0.82\\

SVM\textsubscript{poly} & 0.77 & 0.72 & 0.83 & 0.69 & 0.72 & 0.71 & \textbf{0.78} & \textbf{0.74} & \textbf{0.83} & 0.74 & 0.78 & 0.79 & \textbf{0.75} & 0.70 & 0.81\\

SVM\textsubscript{rad} & \textbf{0.78} & \textbf{0.74} & 0.83 & \textbf{0.70} & \textbf{0.81} & 0.67 & 0.75 & 0.70 & 0.81 & 0.74 & 0.77 & 0.78 & 0.73 & 0.65 & \textbf{0.83}\\
            \bottomrule
        \end{tabular}
        \end{adjustbox}
        \caption{\textbf{F1}, \textbf{P}recision, and \textbf{R}ecall scores for predicting left--right political leaning on the 20\% fully-held out test data (averaged across five balanced samples). Five datasets are used: \textbf{network} friends data only, \textbf{non-political} tweet text, 
        non-political and network data (\textbf{non-pol+net}),
        \textbf{political} tweet text only, and 
        political text and network data (\textbf{pol+net}). Results are shown for five classifiers: Na\"ive Bayes,  Neural Network, and SVMs with \textbf{lin}ear, \textbf{poly}nomial, and \textbf{rad}ial kernels. The highest F1, precision, and recall values for each dataset (column)  are in bold.}
        \label{tbl:classifier_results}
\end{table*}

\subsubsection{Classification procedure}

Our classification procedure follows these steps: (1) We split each sample data into training (80\%) and test (20\%) datasets. (2) We create a topic model using only the training data. (3) We train the political leaning classifiers on the training data using 10-fold cross-validation. (4) We apply the topic model to test data, and (5) we predict the leanings of test data. We completely isolate the test dataset from both the topic generation and classifier training processes by following this procedure.

We use  Structural Topic Models (STM) to create our topic models. However, we do not provide prevalence co-variates to prevent biased topic generation. Therefore, we feed the topic modelling algorithm with only text data. In this case, the algorithm becomes an implementation of the Correlated Topic Model (CTM) \cite{roberts_stm_2019}.

We treat the number of topics as a hyperparameter that we tune through cross-validation on the training data. We find 150 topics 
gives the most meaningful results due to the large corpora (3.8M) and long-period (approx. nine years). We use spectral initialization as advised by the creators since it generates `better' and more `consistent' topics by using ``a spectral decomposition (non-negative matrix factorization) of the word co-occurrence matrix" \cite{roberts_stm_2019}. We extract the theta distributions to classify documents (users) when the model returns. Theta states the probability of a document belonging to a topic.
    
To infer political leanings and compare the results, we train and compare neural network (NN), Na\"ive Bayes (NB), and support-vector machine (SVM) classifiers. The SVMs are trained with linear, radial and polynomial kernels. 
SVM, NB and NN are three of the most popular machine learning methods to classify text, although the `best' approach depends heavily on the data and the specific task being performed \cite{wang_baselines_2012,ng_discriminative_2001}.

After performing a classification, we can apply a threshold to the probabilistic results from the classifier. 
If the probability of a user's classification is above the threshold, we assign them the predicted outcome. When the probability is below the threshold, we assign a value of `unknown' for each best-performing model in Table  \ref{tbl:best_model_results}.
The `Unknown' column reports the percentage of users in each classification task with probabilities below the stated threshold.

\subsection{Classification results}
We first calculate F1 scores using 10-fold cross-validation on the training data (Table~\ref{tbl:auc_results}). For those results, we renew the classification with five different balanced, random samples and report the average final F1 scores across all folds and samples. We also report F1, precision, and recall scores on the entirely held out test data (Table~\ref{tbl:classifier_results}).This test data had no role in creating the topic models or other features and better approximates real-world performance.

In Table \ref{tbl:auc_results} we compare the performance of our models on the training data. The neural network, SVM\textsubscript{poly} and SVM\textsubscript{rad} models perform similar to each other for the network dataset. The SVM\textsubscript{lin} and neural network model performances are close on political text and network only datasets. 
The F1 scores for the dataset combining political text and network data are generally lower than for either alone, suggesting the signals from these two sources somewhat overlap.

In Table \ref{tbl:classifier_results}, we compare the neural network, SVM and NB model performances on the unseen test data averaged across five balanced samples. The results show the best performing classifier changes with the task. The neural network model performs best on the political dataset, but the  SVM\textsubscript{poly} classifier has the highest F1 score for both hybrid datasets. Lastly, for the non-political and network datasets the SVM\textsubscript{rad} model has the highest F1 score.

Our main objective in this paper is to understand how well political leaning can be predicted from the non-political text; so, we further explore how different probability thresholds can be applied to the best performing classifiers' outputs and report these in Table \ref{tbl:best_model_results}. 

For the non-political dataset we select the SVM\textsubscript{rad} model, and for the non-political+network dataset we select the SVM\textsubscript{poly} model.
Classifying political leaning from non-political tweets is more challenging than using political tweets. Incorporating network data helps increase coverage and accuracy in the non-political case. However, even without network data, it is possible to infer a political leaning for nearly half of the users with high performance (For a probability threshold of $0.51$, the F1 score is $0.9$, but 56\% of users are classified as being of an unknown political leaning).

Estimating political leaning from political tweets has a similar F1 score to using network data alone. However, combining network and political text data results in lower confidence estimates (and a higher number of unknown users than either text or network data alone at each threshold).

The effect of the probability threshold for models
makes the comparison easier. This threshold excludes users with results of lower certainty and marks them as unknown. The impact on the performance of each threshold is reasonably consistent across input datasets: F1 scores increase with higher thresholds, but so too do the percentages of users classified with an unknown political leaning.  For example, for the non-political input dataset, a 0.62 probability threshold increases the F1 scores by up to 23 percentage points but also marks nearly 97\% of users in the dataset as unknown. The effects are similar for other input datasets but less extreme.

\begin{table}[tb]
\centering
        \begin{tabular}{lccccc}
        \toprule
            Dataset 
            &Prob. Threshold
            &F1
            & Prec.
            &Rec.
            & Unknown
            \\
            \midrule
            net 
             & 0.50 &  0.86 &  0.85 &  0.87 &  0.00 \\
            \hspace{0.5em}(SVM\textsubscript{rad}) 
            & 0.64 &  0.90 &  0.88 &  0.92 &  0.23 \\
             & 0.68 &  0.95 &  0.95 &  0.95 &  0.41  \\
            \addlinespace
       
            non-pol
             & 0.50 &  0.77 &  0.75 &  0.79 &  0.00  \\
             \hspace{0.5em}(SVM\textsubscript{rad}) 
             & 0.51 &  0.90 &  0.93 &  0.87 &  0.56  \\
             & 0.62 &  1.00 &  1.00 &  1.00 &  0.97 \\
            \addlinespace
       
            non-pol+net
             & 0.50 &  0.85 &  0.82 &  0.88 &  0.00 \\ 
             \hspace{0.5em}(SVM\textsubscript{poly})
             & 0.70 &  0.90 &  0.90 &  0.90 &  0.36\\
             & 0.78 &  0.95 &  0.95 &  0.95 &  0.47 \\
            \addlinespace 
        
            pol 
             & 0.50 &  0.85 &  0.83 &  0.87 &  0.00   \\
             \hspace{0.5em}(NN)
             & 0.84 &  0.90 &  0.88 &  0.92 &  0.27 \\
             & 0.94 &  0.95 &  0.92 &  0.97 &  0.41  \\
            \addlinespace
       
            pol+net
             & 0.50 &  0.81 &  0.75 &  0.87 &  0.00  \\
             \hspace{0.5em}(SVM\textsubscript{poly})
             & 0.74 &  0.91 &  0.90 &  0.92 &  0.52  \\
             & 0.76 &  0.96 &  0.97 &  0.94 &  0.59 \\
            \bottomrule\\
        \end{tabular}
        
        \caption{Example thresholds and F1, precision, and recall scores for the best performing models (the models with the highest F1 scores in Table \protect\ref{tbl:classifier_results}) for each dataset.  
        Different probability thresholds are set to give base, 0.9 and 0.95 F1 scores.}
        \label{tbl:best_model_results}
\end{table}

\subsubsection{Post-hoc analysis} We extract the eleven most important topics and the ten most influential Twitter accounts from our best performing non-political+network SVM\textsubscript{poly} model. 

Table \ref{tbl:imp_network} shows the 10 most significant Twitter accounts in the SVM\textsubscript{poly} non-political+network model. We generate the ratio of the normalized number of left and right followers of these accounts using our ground-truth dataset, and find a higher proportion of users following @10DowningStreet, @realDonaldTrump, and @JeremyClarkson are right-leaning. The remaining accounts were followed more by left-leaning users than right-leaning users. Although most of these accounts are political, there are two non-political accounts: Jeremy Clarkson---the former Top Gear presenter---and Charlton Brooker---the creator of Black Mirror.

\begin{table}[tb]
\centering
        \begin{tabular}{llcc}
        \toprule
            Account 
            & Screen Name
            & Left (\%)
            & Right (\%)
            \\
            \midrule
            Jeremy Corbyn&@jeremycorbyn&\textbf{47}&11 \\
       
            Owen Jones&@OwenJones84&\textbf{33}&4 \\
           
            UK Prime Minister&@10DowningStreet&17&\textbf{36} \\
           
            Donald J. Trump&@realDonaldTrump&16&\textbf{39} \\
            
            Charlie Brooker&@charltonbrooker&\textbf{30}&8 \\
           
            Caroline Lucas&@CarolineLucas&\textbf{25}&2 \\
          
            Jeremy Clarkson&@JeremyClarkson&15&\textbf{35} \\
          
            The Green Party&@TheGreenParty&\textbf{20}&3 \\
          
            Jon Snow&@jonsnowC4&\textbf{25}&7 \\
           
            NHS Million&@NHSMillion&\textbf{21}&3 \\
            \bottomrule\\
        \end{tabular}
        \caption{Most influential Twitter accounts, which are generated by the non-political+network model, sorted by their importance.
        The Left column displays the percentage of users in the groundtruth data labelled as left-leaning who follow the account. The Right column displays the equivalent value for right-leaning users.
        All ten of the accounts are verified (have a blue checkmark badge) on Twitter.}
        \label{tbl:imp_network}
\end{table}

\begin{figure}
\begin{center}
\includegraphics{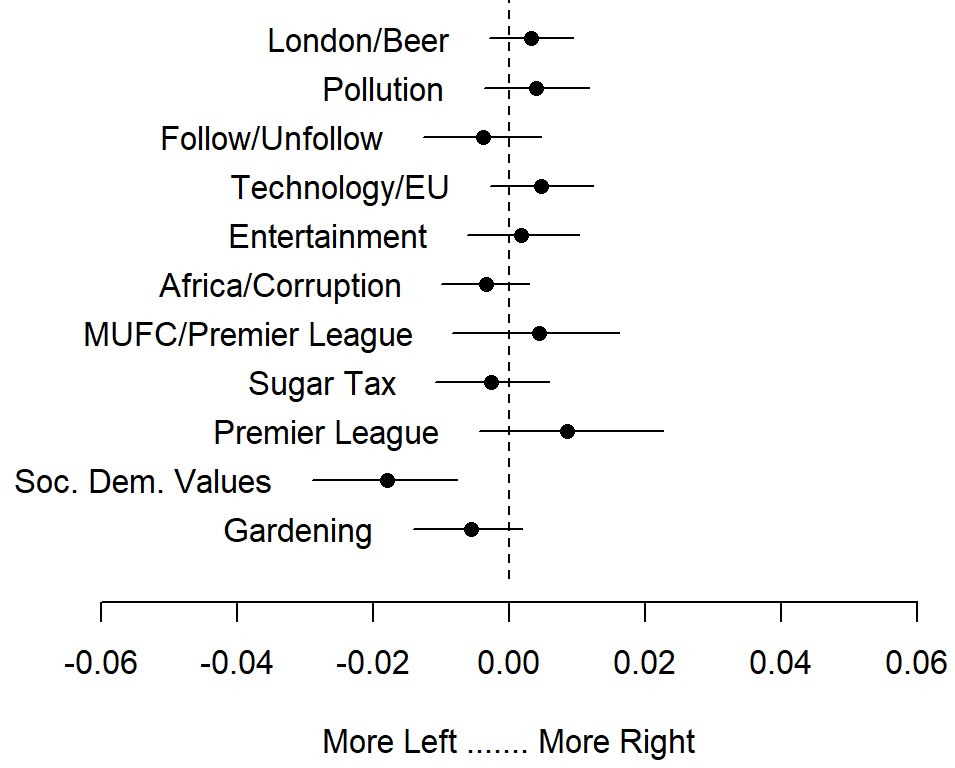}    
\caption{
Topic prevalence. We estimate a regression predicting the proportion of each document about each topic in the STM model using political leaning. We take a global approach in approximating the measurement uncertainty in the topic proportions. We report the topic proportion estimates and 95\% confidence intervals for the 11 topics with the largest left--right differences. For more details see~\protect\citet{roberts_stm_2019}.
}
\label{fig:topic_prevalence}
\end{center}
\end{figure}

After detecting the most important topics, we assign a label to each based on the most prevalent words and estimate the effects of left-leaning and right-leaning users.
Figure \ref{fig:topic_prevalence} shows the contrast between leanings. The top words belonging to these topics are displayed in Table \ref{tbl:topic_words} and a complete list of all topics is available in the supplemental materials. The word lists are developed in two steps. First, we list the top 10 words by using FREX, LIFT and log scores. Then we concatenate these three lists in the given order and extract the top 15 unique words.\footnote{``FREX is the weighted harmonic mean of the word's rank in terms of exclusivity and frequency,'' ``LIFT weights words by dividing by their frequency in other topics, therefore giving higher weight to words that appear less frequently in other topics,'' and ``Score divides the log frequency of the word in the topic by the log frequency of the word in other topics'' \cite{roberts_stm_2019}}.

Figure \ref{fig:topic_prevalence} shows topics discussing social democratic values are more prevalent in left-leaning users. A gardening related topic is also more left-leaning while the Premier League and a pollution-related topic are more right-leaning. It should be noted that most 95\% confidence intervals cross the zero point indicating any left--right distinction is subtle.

\begin{table}[tb]
\centering
    \begin{tabular}{p{2cm}p{7cm}} 
    \toprule
         \textbf{Topic Name} & \textbf{Features}
         \\ \midrule
        London\slash{}Beer & unit\_kingdom, kingdom, ben, unit, beer, pub, london, januari, station, tim, wimbledon, cheat, rip, slice, peak
        \\\addlinespace
 
        Pollution & plastic, contain, ocean, pollut, dr, usa, sea, recycl, dwp, path, locat, commit, toe, current, planet
        \\\addlinespace
    
        Follow\slash{}Unfollow & unfollow, automat, peopl\_follow, stat, follow, check, person, ireland, mention, reach, irish, track, retweet, big\_fan, joe
        \\\addlinespace
      
        Technology\slash{}EU & android, dj, ee, eu, app, greec, independ, mobil, scotland, ipad, euro, europ, io, beat, scottish
        \\\addlinespace
    
        Entertainment & beth, xx, haha, xxx, deploy, sean, artist, nicki, wire, hahaha, jay, spain, boyfriend, til, willi
        \\\addlinespace
   
        Africa\slash{}Corruption & ni, nigeria, lawyer, african, corrupt, parcel, bbcradio, lol, journo, arrest, wk, individu, brazil, pregnant, id
        \\\addlinespace
       
        MUFC\slash{}Premier League & mufc, manutd, unit, rooney, manchest\_unit, utd, goal, golf, moy, mourinho, ronaldo, pogba, score, fifa, leagu
        \\\addlinespace
  
        Sugar Tax & frog, level, skill, c***, sugar, committe, divin, nhs, suspect, holland, hunter, jeremi, bullshit, jeremi\_hunt, ww
        \\\addlinespace
  
        Premier League & west\_ham, spur, ham, salah, everton, kane, chelsea, klopp, pogba, lukaku, goal, mourinho, tottenham, midfield, liverpool
        \\\addlinespace
 
        Soc.\ Dem.\ Values & foodbank, poverti, nhs, wage, worker, crisi, fund, wick, solidar, food\_bank, grenfel, bank, auster, polit, privat
        \\\addlinespace

        Gardening & bloom, garden, cute, layer, kinda, ship, charact, rich, bless, hug, glad, mom, damn, lmao, gonna
        \\
        \bottomrule\\
    \end{tabular}
    \caption{Words belonging to corresponding topics. Unique words are selected by combining top 10 words of FREX, LIFT and log scores in that order. Full topic models are in the supplemental materials.}
    \label{tbl:topic_words}
\end{table}

\subsubsection{Effect of political activity}
After labelling political tweets as discussed, we develop a political activity index by dividing the number of political tweets by the number of overall tweets per user. To determine whether there is a relation between political activity and classifier performance, we apply a Pearson correlation test to the classifier probabilities and our measure of political activity. For the best performing non-political textual model (SVM\textsubscript{rad}), we find only very weak correlations between the classifier probability 
and 
political activity ($r=0.15$), the number of tokens ($r=0.02$), and leaning scores ($r=-0.02$).
The best-performing political textual model (NN) similarly has weak correlations between its predicted probabilities and political activity ($r=0.21$), the number of tokens ($r=0.17$), and leaning scores ($r=0.13$).

\section{Case study: Polarization and news sharing}\label{sec:news_sharing}

Having developed a classifier that successfully predicts the users left--right political leaning even with non-political text as input we now seek to apply this classifier in a short case study.
Social media has become intertwined with ideas of political polarization and echo chambers in popular culture~\cite{pariser2011}. There are, of course, many studies that show polarization in explicitly political contexts, such as the hyperlinks of political bloggers~\cite{adamic_political_2005}, 
the sharing of political news~\cite{flaxman2016filter}, and
Twitter activity in the context of politics~\cite{barbera_birds_2015,hong_political_2016}. 

Despite this, there is a lack of studies examining the polarization and echo chamber hypotheses in diverse 
contexts \cite[for more, see,][]{dubois_echo_2018}.
The Internet offers the potential to access various information sources, and enthusiasm for politics and consumption of more media sources negatively coordinate with the likelihood of being in an echo chamber
\cite{dubois_echo_2018, barbera_tweeting_2015}.

In this case study, we apply our non-political classifier to the sharing of news from three UK media outlets: The Telegraph (right-leaning), the BBC (centrist), and The Guardian (left-leaning). The case study fulfils two objectives. First, it demonstrates the face validity of our classifier by matching common expectations that political news from The Telegraph should be shared chiefly by right-leaning users, political news from The Guardian should be shared mostly by left-leaning users, and the BBC should be the least polarized of the three. Second, the case study advances the scholarly conversation on polarization by examining the sharing of non-political news---namely sports news---from these three outlets. Our results show that many right-leaning users consume sports news from The Guardian, suggesting more right-leaning users engage with, at least the sports section of, The Guardian than conventional wisdom would expect.

\subsection{Data and methods}

We examine the number of left- and right-leaning accounts sharing news in a two-by-three design. Our three news sources are The Telegraph, the BBC, and The Guardian, and our new types are political and sports news.

We first use elevated Twitter Streaming API access to collect URL patterns related to 
\url{theguardian.com}, \url{telegraph.co.uk}, and \url{bbc.co.uk}. We then manually examine these URLs to identify URLs patterns that are explicitly associated with \textit{political} news and \textit{sport} news. For example, all URLs containing \url{bbc.co.uk/sport/} are labelled as sports news from the BBC, and all URLs containing \url{theguardian.com/politics/2018/} are labelled as political news from The Guardian. A complete list of URL patterns is available in the supplemental materials.

Returning to the Twitter Streaming API, we collect all instances of URLs shared matching any of our patterns between 1 September 2018 and 8 September 2018. We then 
query the 3,200 most recent tweets of each user and collect the friends of each user appearing in our dataset using the Twitter RESTful API \texttt{statuses/user\_timeline} endpoint and \texttt{friends/ids} endpoint.

Each user's recent tweets and friends are used as input to estimate each user's left--right political leaning. Next, we clean the text as described in the development of our classifier and combine all tweets from a given user into one document per user to estimate the political leaning of that user. The DFMs result in extremely sparse matrices (e.g., approx.\ 25M features). We, therefore, remove terms whose total frequency is lower than three, which makes the DFM dimensions
computationally-feasible 
(e.g., approx.\ 5M features). Then, we extract features which are not present in our non-political DFM. As a result, we also remove political terms in this step.  In the last step, we fit the non-political topic model to the news data to predict the political leaning of each user.

Our setup matches the non-political + network approach described in the main article. 
To classify our users sharing political and sports news, we select the SVM\textsubscript{poly} classifier that yielded the optimal 
performance in the non-political + network task and set probability threshold to 0.7.

\subsection{News sharing results}
The results of our classification are shown in Table~\ref{tbl:news_sharing}. The results conform with our hypothesis that the political leaning of users would match the political-leaning of the news publishers for political news sharing. For example, among users sharing political news who were classified as left-leaning, over 75\% shared articles from the left-leaning Guardian. Similarly, over half of the users are classified as right-leaning who shared political news links from the right-leaning Telegraph.

In contrast to politics, we find The Guardian is the most popular source for sports news among users classified as either left- or right-leaning. In general right-leaning users share more sports than left-leaning users as a whole. Even so, the number of right-leaning users sharing sports news from The Telegraph is 8.5 times higher than the number of left-leaning users sharing articles from that source---the largest proportional difference in the dataset. 
We suspect sports news sharing is influenced by availability: The Telegraph has a paywall while the BBC and The Guardian do not. So it makes sense that subscribers paying for The Telegraph would get both political and sports news from it but that non-subscribers would look elsewhere. Many right-leaning users share sports news from the left-leaning, but free-to-access, Guardian suggests factors beyond political leaning such as availability and cost heavily influence the news sources people read.

Overall, the case study shows our classifier's face-validity being able to detect the expected leanings of users sharing political news. At the same time, it suggests that the sharing of sports news is not as polarized and points toward new avenues of study made possible with a classifier that can predict political leaning from non-political text.

\begin{table}[tb]
    \begin{center}
    \begin{tabular}{lrrrr}
    \toprule
  & Guardian & BBC & Telegraph & Total\\
    \midrule
    
    Political \\
    \hspace{3ex} Left &  \textbf{1,223} &  240 &  161 & 1,624\\
    \hspace{3ex} Right &  812 &  377 &  \textbf{1,230} & 2,419\\
    \hspace{3ex} Unknown &  \textbf{650} &  238 &  334 & 3,641\\
            
Sport \\
    \hspace{3ex} Left &  \textbf{314} &  295 &  44 & 653\\
    \hspace{3ex} Right &  \textbf{1,631} &  869 &  378 & 2,878\\
    \hspace{3ex} Unknown &  \textbf{956} &  564 &  115 & 1,635\\
    \bottomrule\\
    \end{tabular}
    \caption{The political orientation of individuals sharing news by source and type. We set the probability threshold to 0.7 which has a 90\% accuracy is our model development (non-political+network). The highest value in each row, excluding the total, is in bold.}
    \label{tbl:news_sharing}
    \end{center}
\end{table}

\section{Discussion}

Polarization is often studied in explicitly political contexts; however, understanding the actual effects of polarization requires researchers to explore non-political contexts. This, in turn, requires the ability to estimate political leaning in these non-political contexts. Building on the concept of `lifestyle politics' \cite{della2015why} this paper has shown it is possible to estimate political leaning from non-political text.

What exactly is non-political text is unclear. Our preliminary analysis found that tweets sent outside of election periods were still often political in nature (e.g., containing discussion about the UK's relationship with Europe). In response, we developed an approach to define and expand a list of political keywords and only allow tweets without these keywords to be considered for our tasks with non-political text. The topics displayed in Table \ref{tbl:topic_words} suggest this worked reasonably well.
However, a qualitative inspection of the topics reveals a small number of political tweets are still in the non-political dataset: eight out of the 150 topics (5.3\%) are political (Israel--Palestine, Scottish independence, Brexit, Macron--Merkel, the Green Party, UK Politics, US Politics, Westminster). Nonetheless, these form a tiny part of the dataset and cannot be responsible for the performance of our classifiers. 
Indeed, applying the best performing classifier trained on political text to the non-political text results in accuracy no better than random chance.

We find only only weak correlations between classifier performance and either political activity or political leaning. This suggests that individuals with different activity levels and at different points in the left--right spectrum all leave traces of their political leaning. 

The use of Voter Advice Applications (VAAs) for ground truth data represents an exciting way to build larger-datasets for training and testing machine learning classifiers. 
In contrast to human coders~\cite[e.g.,][]{conover_predicting_2011} or crowdsourcing \cite[e.g.,][]{iyyer_political_2014}, VAAs are self-reported. \citet{preotiuc-pietro_beyond_2017} used self-report surveys, but VAAs can result in larger sample sizes, provide a longer assessment, and rely on the genuine motivation of users who also learn how their political preferences align with major parties. Nonetheless, it will be important to understand any biases in who shares VAA results on social media platforms. For instance, it may be that only the most politically-engaged users share VAA results on social media.

In addition to the variety of input data used, prior research has also tried various machine learning approaches to classifying political leaning including SVM, Na\"ive Bayes, and neural networks. Reported performance scores range from 60\% to 90\% \cite{preotiuc-pietro_beyond_2017,gu_ideology_2016,jelveh_detecting_2014,iyyer_political_2014,yu_classifying_2008,kulkarni_multi-view_2018}. 
Thus, our F1 score of 0.81 for political text is good, and achieving an F1 score of 0.7 for non-political text alone or 0.75 for non-political text and network data is in line with other work using explicitly political text.

Ideology, of course, is much broader than a simple left--right or liberal--conservative spectrum. Our work, therefore, complements approaches such as \citet{graham_moral_2012} who apply moral foundation theory and find liberals and conservatives do not differ dramatically on moral foundations. The topics we find echo the moral foundations Graham et al. identify: liberals give higher importance to individualizing foundations---harm/care and fairness/reciprocity---and conservatives give higher importance to binding foundations---in-group/loyalty, authority/respect, and purity/sanctity. 
That the topic on pollution is more associated with the right may seem counter-intuitive, but we note the effect sizes are small and the analysis only considers prevalence, not attitude or sentiment.

We find the accounts that a user follows often reveal that user's political leaning. While some of these accounts are explicitly political, others such as Jeremy Clarkson, a former Top Gear presenter, and Charlton Brooker, the creator of Black Mirror, are not. 
In other words, interest in Top Gear/Grand Tour or Black Mirror may hint at individuals' political leaning more than expected. It may be unexpected that an official government account such as that of the UK Prime Minister is so explicitly political, but it is worth noting in the UK context that the Conservative Party has been in power since 2010.

Although our F1 scores indicate there are left/right differences, claiming that all right and left users behave similarly would be a gross oversimplification. There may be significant differences in other countries and cultures.
In addition, our data is not representative. We cannot generalize the results of this study to explain complex structures such as conservative or social-democratic ideologies.

Our brief case study of news sharing hints at promising ways in which classifiers of non-political content may be used. We found that political news sharing on Twitter is associated with the political leaning of users, but that sports news sharing is less polarised. Many accounts classified as right-leaning shared sports news from The Guardian, a left-leaning source. 
Telegraph sports news, however, was still heavily associated with right-leaning users. We suspect paywalls partially drive this result: The Telegraph is the only source with a paywall in our data.
It may be that many right-leaning users look to The Guardian and other sources for sports news rather than pay for access to The Telegraph. On the other hand, individuals who already pay for access to The Telegraph likely also choose to get their sports news there.

The level of polarization observed on social media thus depends in part on the type of content examined. As mentioned, most studies examine polarization in the context of the sharing of explicitly political content. However, our results indicate that a more holistic view of the content shared beyond politics may show less polarization. Our results also suggest that editorial decisions such as having a paywall or not affect the level of polarization observed in link-sharing on social media.

\section{Conclusions}

As political orientation influences and correlates with many aspects of non-political life, there is good reason to expect that political leaning can be inferred even from non-political text.

We first developed a classifier using political text from tweets primarily associated with general elections in the UK, achieving an F1 score of 0.81. 
We then collected recent tweets from users outside of any election period and removed tweets with political keywords. Using this non-political text data to estimate the political leaning of users, we were still able to achieve an F1 score of 0.70. Furthermore, incorporating the network information of these users alongside the textual features from their tweets further increased the F1 score.

Using our classifier, we were able to classify the political leaning of users sharing political and sports news on Twitter from three UK national news sources. Our analysis indicated the importance of examining polarization in political and non-political contexts.

This study has shown the possibility for the classification of political leaning from non-political text and highlighted the importance of studying political leaning in non-political contexts. We hope that this paper can help spur research to understand polarization in non-political contexts.

\bibliographystyle{spbasic} 
\bibliography{main}

\begin{thebibliography}{46}
\providecommand{\natexlab}[1]{#1}
\providecommand{\url}[1]{{#1}}
\providecommand{\urlprefix}{URL }
\expandafter\ifx\csname urlstyle\endcsname\relax
  \providecommand{\doi}[1]{DOI~\discretionary{}{}{}#1}\else
  \providecommand{\doi}{DOI~\discretionary{}{}{}\begingroup
  \urlstyle{rm}\Url}\fi
\providecommand{\eprint}[2][]{\url{#2}}

\bibitem[{Adamic and Glance(2005)}]{adamic_political_2005}
Adamic LA, Glance N (2005) The political blogosphere and the 2004 u.s.
  election: Divided they blog. In: Proceedings of the 3rd International
  Workshop on Link Discovery, ACM, New York, NY, USA, LinkKDD '05, pp 36--43,
  \doi{10.1145/1134271.1134277},
  \urlprefix\url{http://doi.acm.org/10.1145/1134271.1134277}

\bibitem[{Ahn et~al.(2014)Ahn, Kishida, Gu, Lohrenz, Harvey, Alford, Smith,
  Yaffe, Hibbing, Dayan, and Montague}]{ahn_nonpolitical_2014}
Ahn WY, Kishida K, Gu X, Lohrenz T, Harvey A, Alford J, Smith K, Yaffe G,
  Hibbing J, Dayan P, Montague PR (2014) Nonpolitical images evoke neural
  predictors of political ideology 24(22):2693--2699,
  \doi{10.1016/j.cub.2014.09.050},
  \urlprefix\url{http://www.sciencedirect.com/science/article/pii/S0960982214012135}

\bibitem[{Barberá(2015)}]{barbera_birds_2015}
Barberá P (2015) Birds of the {Same} {Feather} {Tweet} {Together}. {Bayesian}
  {Ideal} {Point} {Estimation} {Using} {Twitter} {Data} p~28

\bibitem[{Barberá et~al.(2015)Barberá, Jost, Nagler, Tucker, and
  Bonneau}]{barbera_tweeting_2015}
Barberá P, Jost JT, Nagler J, Tucker JA, Bonneau R (2015) Tweeting from left
  to right: Is online political communication more than an echo chamber?
  26(10):1531--1542, \doi{10.1177/0956797615594620},
  \urlprefix\url{http://journals.sagepub.com/doi/10.1177/0956797615594620}

\bibitem[{Chen et~al.(2015)Chen, Wang, Agichtein, and
  Wang}]{chen_comparative_2015}
Chen X, Wang Y, Agichtein E, Wang F (2015) A comparative study of demographic
  attribute inference in twitter p~4

\bibitem[{Compton et~al.(2014)Compton, Jurgens, and
  Allen}]{compton_geotagging_2014}
Compton R, Jurgens D, Allen D (2014) Geotagging one hundred million twitter
  accounts with total variation minimization. In: 2014 {IEEE} International
  Conference on Big Data (Big Data), {IEEE}, pp 393--401,
  \doi{10.1109/BigData.2014.7004256},
  \urlprefix\url{http://ieeexplore.ieee.org/document/7004256/}

\bibitem[{Conover et~al.(2011)Conover, Goncalves, Ratkiewicz, Flammini, and
  Menczer}]{conover_predicting_2011}
Conover MD, Goncalves B, Ratkiewicz J, Flammini A, Menczer F (2011) Predicting
  the {Political} {Alignment} of {Twitter} {Users}. In: 2011 {IEEE} {Third}
  {International} {Conference} on {Privacy}, {Security}, {Risk} and {Trust} and
  2011 {IEEE} {Third} {International} {Conference} on {Social} {Computing}, pp
  192--199, \doi{10.1109/PASSAT/SocialCom.2011.34}

\bibitem[{DellaPosta et~al.(2015)DellaPosta, Shi, and Macy}]{della2015why}
DellaPosta D, Shi Y, Macy M (2015) Why do liberals drink lattes? American
  Journal of Sociology 120(5):1473--1511, \doi{10.1086/681254},
  \urlprefix\url{https://doi.org/10.1086/681254}

\bibitem[{Diermeier et~al.(2012)Diermeier, Godbout, Yu, and
  Kaufmann}]{diermeier_language_2012}
Diermeier D, Godbout JF, Yu B, Kaufmann S (2012) Language and ideology in
  congress 42(1):31--55, \doi{10.1017/S0007123411000160},
  \urlprefix\url{https://www.cambridge.org/core/journals/british-journal-of-political-science/article/language-and-ideology-in-congress/1063F5509BC2ABC3F9A0E164E58157EE}

\bibitem[{Dubois and Blank(2018)}]{dubois_echo_2018}
Dubois E, Blank G (2018) The echo chamber is overstated: the moderating effect
  of political interest and diverse media 21(5):729--745,
  \doi{10.1080/1369118X.2018.1428656},
  \urlprefix\url{https://doi.org/10.1080/1369118X.2018.1428656}

\bibitem[{Eysenck(1964)}]{eysenck_sense_1964}
Eysenck HJ (1964) Sense and nonsense in psychology. Penguin Books

\bibitem[{Flaxman et~al.(2016)Flaxman, Goel, and Rao}]{flaxman2016filter}
Flaxman S, Goel S, Rao JM (2016) Filter bubbles, echo chambers, and online news
  consumption. Public Opinion Quarterly 80(S1):298--320,
  \doi{10.1093/poq/nfw006}, \urlprefix\url{https://doi.org/10.1093/poq/nfw006}

\bibitem[{Golbeck and Hansen(2014)}]{golbeck_method_2014}
Golbeck J, Hansen D (2014) A method for computing political preference among
  {Twitter} followers. Social Networks 36:177--184,
  \doi{10.1016/j.socnet.2013.07.004},
  \urlprefix\url{http://www.sciencedirect.com/science/article/pii/S0378873313000683}

\bibitem[{Graham et~al.(2012)Graham, Nosek, and Haidt}]{graham_moral_2012}
Graham J, Nosek BA, Haidt J (2012) The moral stereotypes of liberals and
  conservatives: Exaggeration of differences across the political spectrum.
  {PLoS} {ONE} 7(12):e50092, \doi{10.1371/journal.pone.0050092},
  \urlprefix\url{https://dx.plos.org/10.1371/journal.pone.0050092}

\bibitem[{Gu et~al.(2016)Gu, Chen, Sun, and Wang}]{gu_ideology_2016}
Gu Y, Chen T, Sun Y, Wang B (2016) Ideology {Detection} for {Twitter} {Users}
  with {Heterogeneous} {Types} of {Links}. arXiv:161208207 [cs]
  \urlprefix\url{http://arxiv.org/abs/1612.08207}, arXiv: 1612.08207

\bibitem[{Hale(2014)}]{hale_global_2014}
Hale SA (2014) Global connectivity and multilinguals in the twitter network.
  In: Proceedings of the 32nd annual {ACM} conference on Human factors in
  computing systems - {CHI} '14, {ACM} Press, pp 833--842,
  \doi{10.1145/2556288.2557203},
  \urlprefix\url{http://dl.acm.org/citation.cfm?doid=2556288.2557203}

\bibitem[{Hong and Kim(2016)}]{hong_political_2016}
Hong S, Kim SH (2016) Political polarization on twitter: Implications for the
  use of social media in digital governments 33(4):777--782

\bibitem[{Iyyer et~al.(2014)Iyyer, Enns, Boyd-Graber, and
  Resnik}]{iyyer_political_2014}
Iyyer M, Enns P, Boyd-Graber J, Resnik P (2014) Political ideology detection
  using recursive neural networks. In: Proceedings of the 52nd Annual Meeting
  of the Association for Computational Linguistics (Volume 1: Long Papers),
  Association for Computational Linguistics, pp 1113--1122,
  \doi{10.3115/v1/P14-1105},
  \urlprefix\url{http://aclweb.org/anthology/P14-1105}

\bibitem[{Jelveh et~al.(2014)Jelveh, Kogut, and Naidu}]{jelveh_detecting_2014}
Jelveh Z, Kogut B, Naidu S (2014) Detecting latent ideology in expert text:
  Evidence from academic papers in economics. In: Proceedings of the 2014
  Conference on Empirical Methods in Natural Language Processing ({EMNLP}),
  Association for Computational Linguistics, pp 1804--1809,
  \doi{10.3115/v1/D14-1191},
  \urlprefix\url{http://aclweb.org/anthology/D14-1191}

\bibitem[{King et~al.(2016)King, Orlando, and Sparks}]{king_ideological_2016}
King AS, Orlando FJ, Sparks DB (2016) Ideological {Extremity} and {Success} in
  {Primary} {Elections}: {Drawing} {Inferences} {From} the {Twitter} {Network}.
  Social Science Computer Review 34(4):395--415,
  \doi{10.1177/0894439315595483},
  \urlprefix\url{https://doi.org/10.1177/0894439315595483}

\bibitem[{Kitschelt(1994)}]{kitschelt_transformation_1994}
Kitschelt H (1994) The Transformation of European Social Democracy by Herbert
  Kitschelt. \doi{10.1017/CBO9780511622014},
  \urlprefix\url{/core/books/transformation-of-european-social-democracy/C92F284FC17302253C3B5B14123BBA80}

\bibitem[{Kulkarni et~al.(2018)Kulkarni, Ye, Skiena, and
  Wang}]{kulkarni_multi-view_2018}
Kulkarni V, Ye J, Skiena S, Wang WY (2018) Multi-view models for political
  ideology detection of news articles
  \urlprefix\url{http://arxiv.org/abs/1809.03485}, \eprint{1809.03485}

\bibitem[{Lewis et~al.(2004)Lewis, Yang, Rose, and Li}]{lewis_rcv1:_2004}
Lewis DD, Yang Y, Rose TG, Li F (2004) {RCV}1: A new benchmark collection for
  text categorization research pp 361--397

\bibitem[{Liu et~al.(2019)Liu, Cheng, Pei, Shu, Ge, Ma, Du, Ou, Wang, and
  Xu}]{liu_inferring_2019}
Liu Y, Cheng D, Pei T, Shu H, Ge X, Ma T, Du Y, Ou Y, Wang M, Xu L (2019)
  Inferring gender and age of customers in shopping malls via indoor
  positioning data p 2399808319841910, \doi{10.1177/2399808319841910},
  \urlprefix\url{https://doi.org/10.1177/2399808319841910}

\bibitem[{Ng and Jordan(2001)}]{ng_discriminative_2001}
Ng AY, Jordan MI (2001) On discriminative vs. generative classifiers: A
  comparison of logistic regression and naive bayes. In: Proceedings of the
  14th International Conference on Neural Information Processing Systems:
  Natural and Synthetic, {MIT} Press, {NIPS}'01, pp 841--848,
  \urlprefix\url{http://dl.acm.org/citation.cfm?id=2980539.2980648},
  event-place: Vancouver, British Columbia, Canada

\bibitem[{Nguyen et~al.(2013)Nguyen, Gravel, Trieschnigg, and
  Meder}]{nguyen_how_2013}
Nguyen D, Gravel R, Trieschnigg D, Meder T (2013) "how old do you think i am?"
  a study of language and age in twitter. In: {ICWSM}

\bibitem[{Nguyen et~al.(2016)Nguyen, Doğruöz, Rosé, and
  de~Jong}]{nguyen2016computational}
Nguyen D, Doğruöz AS, Rosé CP, de~Jong F (2016) Computational
  sociolinguistics: A survey. Computational Linguistics 42(3):537--593,
  \doi{10.1162/COLI\_a\_00258},
  \urlprefix\url{https://doi.org/10.1162/COLI_a_00258}

\bibitem[{Pariser(2011)}]{pariser2011}
Pariser E (2011) {The filter bubble: What the {Internet} is hiding from you}.
  Viking, London

\bibitem[{Pennacchiotti and Popescu(2011)}]{pennacchiotti_democrats_2011}
Pennacchiotti M, Popescu AM (2011) Democrats, republicans and starbucks
  afficionados: user classification in twitter p~9

\bibitem[{Porter(1980)}]{porter_algorithm_1980}
Porter MF (1980) An algorithm for suffix stripping 40:211--218,
  \doi{10.1108/00330330610681286}

\bibitem[{Preoţiuc-Pietro et~al.(2017)Preoţiuc-Pietro, Liu, Hopkins, and
  Ungar}]{preotiuc-pietro_beyond_2017}
Preoţiuc-Pietro D, Liu Y, Hopkins D, Ungar L (2017) Beyond {Binary} {Labels}:
  {Political} {Ideology} {Prediction} of {Twitter} {Users}. In: Proceedings of
  the 55th {Annual} {Meeting} of the {Association} for {Computational}
  {Linguistics} ({Volume} 1: {Long} {Papers}), Association for Computational
  Linguistics, Vancouver, Canada, pp 729--740, \doi{10.18653/v1/P17-1068},
  \urlprefix\url{http://aclweb.org/anthology/P17-1068}

\bibitem[{Rao et~al.(2010)Rao, Yarowsky, Shreevats, and
  Gupta}]{rao_classifying_2010}
Rao D, Yarowsky D, Shreevats A, Gupta M (2010) Classifying latent user
  attributes in {T}witter. In: Proceedings of the 2nd international workshop on
  {Search} and mining user-generated contents - {SMUC} '10, ACM Press, Toronto,
  ON, Canada, p~37, \doi{10.1145/1871985.1871993},
  \urlprefix\url{http://portal.acm.org/citation.cfm?doid=1871985.1871993}

\bibitem[{Roberts et~al.(2019)Roberts, Stewart, and Tingley}]{roberts_stm_2019}
Roberts ME, Stewart BM, Tingley D (2019) stm: An {R} package for structural
  topic models 91(1):1--40, \doi{10.18637/jss.v091.i02},
  \urlprefix\url{https://www.jstatsoft.org/index.php/jss/article/view/v091i02},
  number: 1

\bibitem[{Rosenthal and {McKeown}(2011)}]{rosenthal_age_2011}
Rosenthal S, {McKeown} K (2011) Age prediction in blogs: A study of style,
  content, and online behavior in pre- and post-social media generations p~10

\bibitem[{Sap et~al.(2014)Sap, Park, Eichstaedt, Kern, Stillwell, Kosinski,
  Ungar, and Schwartz}]{sap_developing_2014}
Sap M, Park G, Eichstaedt J, Kern M, Stillwell D, Kosinski M, Ungar L, Schwartz
  HA (2014) Developing age and gender predictive lexica over social media. In:
  Proceedings of the 2014 Conference on Empirical Methods in Natural Language
  Processing ({EMNLP}), Association for Computational Linguistics, pp
  1146--1151, \doi{10.3115/v1/D14-1121},
  \urlprefix\url{http://aclweb.org/anthology/D14-1121}

\bibitem[{Sites(2013)}]{cld2}
Sites D (2013) Compact language detector 2.
  \urlprefix\url{https://github.com/CLD2Owners/cld2}

\bibitem[{Sznajd-Weron and Sznajd(2005)}]{sznajd-weron_who_2005}
Sznajd-Weron K, Sznajd J (2005) Who is left, who is right? 351(2):593--604,
  \doi{10.1016/j.physa.2004.12.038},
  \urlprefix\url{http://www.sciencedirect.com/science/article/pii/S0378437104016061}

\bibitem[{Talaifar and Swann(2019)}]{talaifar_deep_2019}
Talaifar S, Swann WB (2019) Deep alignment with country shrinks the moral gap
  between conservatives and liberals 40(3):657--675, \doi{10.1111/pops.12534},
  \urlprefix\url{https://onlinelibrary.wiley.com/doi/abs/10.1111/pops.12534},
  \_eprint: https://onlinelibrary.wiley.com/doi/pdf/10.1111/pops.12534

\bibitem[{Wang and Manning(2012)}]{wang_baselines_2012}
Wang S, Manning CD (2012) Baselines and bigrams: Simple, good sentiment and
  topic classification. In: Proceedings of the 50th Annual Meeting of the
  Association for Computational Linguistics: Short Papers - Volume 2,
  Association for Computational Linguistics, {ACL} '12, pp 90--94,
  \urlprefix\url{http://dl.acm.org/citation.cfm?id=2390665.2390688},
  event-place: Jeju Island, Korea

\bibitem[{Wang et~al.(2019)Wang, Hale, Adelani, Grabowicz, Hartman,
  {FlÃ}\{{\textbackslash}textbackslash\}Pck, and
  Jurgens}]{wang_demographic_2019}
Wang Z, Hale S, Adelani DI, Grabowicz P, Hartman T,
  {FlÃ}\{{\textbackslash}textbackslash\}Pck F, Jurgens D (2019) Demographic
  inference and representative population estimates from multilingual social
  media data. In: The World Wide Web Conference, {ACM}, {WWW} '19, pp
  2056--2067, \doi{10.1145/3308558.3313684},
  \urlprefix\url{http://doi.acm.org/10.1145/3308558.3313684}, event-place: San
  Francisco, {CA}, {USA}

\bibitem[{Weber et~al.(2013)Weber, Garimella, and Teka}]{Weber2013PoliticalHT}
Weber I, Garimella VRK, Teka A (2013) Political hashtag trends. In: ECIR

\bibitem[{Wong et~al.(2016)Wong, Tan, Sen, and Chiang}]{wong_quantifying_2016}
Wong FMF, Tan CW, Sen S, Chiang M (2016) Quantifying {Political} {Leaning} from
  {Tweets}, {Retweets}, and {Retweeters}. IEEE Transactions on Knowledge and
  Data Engineering 28(8):2158--2172, \doi{10.1109/TKDE.2016.2553667}

\bibitem[{Yu et~al.(2008)Yu, Kaufmann, and Diermeier}]{yu_classifying_2008}
Yu B, Kaufmann S, Diermeier D (2008) Classifying party affiliation from
  political speech 5(1):33--48, \doi{10.1080/19331680802149608},
  \urlprefix\url{https://doi.org/10.1080/19331680802149608}

\bibitem[{Zagheni et~al.(2017)Zagheni, Weber, and
  Gummadi}]{zagheni_leveraging_2017}
Zagheni E, Weber I, Gummadi K (2017) Leveraging facebook's advertising platform
  to monitor stocks of migrants 43(4):721--734, \doi{10.1111/padr.12102},
  \urlprefix\url{https://onlinelibrary.wiley.com/doi/abs/10.1111/padr.12102}

\bibitem[{Zamal et~al.(2012)Zamal, Liu, and Ruths}]{zamal_homophily_2012}
Zamal FA, Liu W, Ruths D (2012) Homophily and {Latent} {Attribute} {Inference}:
  {Inferring} {Latent} {Attributes} of {Twitter} {Users} from {Neighbors}. pp
  1--6, \doi{10.1109/IWBF.2017.7935106}

\bibitem[{Zhang et~al.(2016)Zhang, Hu, Zhang, and Liu}]{zhang_your_2016}
Zhang J, Hu X, Zhang Y, Liu H (2016) Your age is no secret: Inferring
  microbloggers' ages via content and interaction analysis p~10

\end{thebibliography}

\appendix






\section{Supplemental materials}

\noindent This supplemental material includes additional detail on how we identify political content, the URL patterns we use to identify the sharing of political and sport news, and all topics in the topic model for non-political and also political content using 150 topics.

\subsection{Identifying political content}
\subsubsection{Political Words}

\seqsplit{
    \#askmay, \#battlefornumber10, \#bbcdebate, \#bbcelection, \#bbcqt, \#bbcsp, \#believeinbritain, \#brexit, \#brexitdeal, \#bristolgreenmp, \#bristolwest, \#budget2015, \#budget2017, \#cameronmustgo, \#conservative, \#conservatives, \#corbyn, \#cpc17, \#dementiatax, \#dupcoalition, \#ed4pm, \#election, \#election2015, \#election2017, \#electionday, \#electionday2017, \#forthemany, \#ge15, \#ge17, \#ge2015, \#ge2017, \#generalelection, \#generalelection17, \#generalelection2017, \#getcameronout, \#greens, \#greensurge, \#hungparliament, \#imvotinglabour, \#itvdebate, \#jc4pm, \#jc4pm2019, \#jeremycorbyn, \#jezwecan, \#labour, \#labourdoorstep, \#labourleadership, \#labourmanifesto, \#labourmustwin, \#le2017, \#leadersdebate, \#libdem, \#libdemfightback, \#libdems, \#makejunetheendofmay, \#marrshow, \#mayvcorbyn, \#newsnight, \#nhscyberattack, \#nigelfarage, \#peston, \#plaid15, \#pmqs, \#queensspeech, \#registertovote, \#ridge, \#saveilf, \#snp, \#snpout, \#tories, \#toriesout, \#toriesoutnow, \#tory, \#toryelectionfraud, \#torymanifesto, \#ukip, \#uklabour, \#victorialive, \#vote, \#vote2017, \#votecameronout, \#voteconservative, \#votegreen2015, \#votegreen2017, \#votelabour, \#votelibdem, \#votematch, \#votesnp, \#voteukip, \#weakandwobbly, \#whyimvotingukip, \#whyvote, \#wsyvf, @aaronbastani, @ainemichellel, @amberruddhr, @amelia\_womack, @andrewspoooner, @andysearson, @angelarayner, @angiemeader, @angry\_voice, @angrysalmond, @annaturley, @barrygardiner, @bbcbreaking, @bbclaurak, @bbcnews, @bbcpolitics, @bbcr4today, @bbcscotlandnews, @bonn1egreer, @borisjohnson, @brexitbin, @bristolgreen, @britainelects, @campbellclaret, @carolinelucas, @cchqpress, @charliewoof81, @christinasnp, @chukaumunna, @chunkymark, @conservatives, @corbyn\_power, @corbynator2, @d\_raval, @daily\_politics, @darrenhall2015, @david\_cameron, @davidjfhalliday, @davidjo52951945, @davidlammy, @davidschneider, @dawnhfoster, @debbie\_abrahams, @dmreporter, @dpjhodges, @dvatw, @ed\_miliband, @el4jc, @emilythornberry, @evolvepolitics, @faisalislam, @fight4uk, @frasernelson, @gdnpolitics, @georgeaylett, @georgeeaton, @grantshapps, @guardian, @guardiannews, @guidofawkes, @hackneyabbott, @harrietharman, @harryslaststand, @hephaestus7, @hrtbps, @huffpostuk, @huffpostukpol, @iainmartin1, @iandunt, @imajsaclaimant, @independent, @ipsosmori, @itvnews, @jacob\_rees\_mogg, @james4labour, @jameskelly, @jamesmelville, @jamieross7, @jeremy\_hunt, @jeremycorbyn, @jeremycorbyn4pm, @jimwaterson, @joglasg, @johnmcdonnellmp, @johnprescott, @johnrentoul, @jolyonmaugham, @jon\_swindon, @jonashworth, @joswinson, @keir\_starmer, @kevin\_maguire, @kezdugdale, @krishgm, @laboureoin, @labourleft, @labourlewis, @labourpress, @ladydurrant, @leannewood, @liamyoung, @libdempress, @libdems, @louisemensch, @lucympowell, @mancman10, @marcherlord1, @markfergusonuk, @mhairihunter, @michaelrosenyes, @mikegalsworthy, @mirrorpolitics, @mmaher70, @molly4bristol, @mollymep, @montie, @mrmalky, @msmithsonpb, @natalieben, @newsthump, @nhaparty, @nhsmillion, @nick\_clegg, @nickreeves9876, @nicolasturgeon, @nigel\_farage, @normanlamb, @nw\_nicholas, @owenjones84, @patricianpino, @paulmasonnews, @paulnuttallukip, @paulwaugh, @peoplesmomentum, @peston, @peterstefanovi2, @petewishart, @plaid\_cymru, @politicshome, @prisonplanet, @rachael\_swindon, @realdonaldtrump, @reclaimthenews, @redhotsquirrel, @redpeter99, @rhonddabryant, @richardburgon, @richardjmurphy, @robmcd85, @rosscolquhoun, @ruthdavidsonmsp, @sarahchampionmp, @scottishlabour, @scottories, @screwlabour, @skwawkbox, @skynews, @skynewsbreak, @socialistvoice, @standardnews, @stvnews, @sunny\_hundal, @survation, @suzanneevans1, @telegraph, @telegraphnews, @telepolitics, @thecanarysays, @thegreenparty, @themingford, @thepileus, @theredrag, @theresa\_may, @theresa\_may's, @therightarticle, @thesnp, @thisisamy\_, @timfarron, @timothy\_stanley, @tnewtondunn, @tomlondon6, @toryfibs, @trevdick, @trobinsonnewera, @uk\_rants, @ukip, @uklabour, @vincecable, @wesstreeting, @willblackwriter, @wingsscotland, @wowpetition, @yougov, @yvettecoopermp, abbot, abbott, bennett, bernie, boris, brexit, cameron, cameron's, campaign, campaigning, candidate, candidates, canvassing, centrist, clegg, clegg's, clinton, coalition, communist, comres, con, conservative, conservatives, constituency, constituents, corbyn, corbyn's, councillor, councillors, cymru, davidson, debates, dem, democrat, democrats, dems, dup, election, elections, electoral, entitlement, eu, fallon, farage, farage's, farron, fein, fiscal, gardiner, ge, gop, gove, govern, government, government's, govt, govt's, greens, grn, guardian, hamas, hammond's, hillary, hustings, icm, ifs, ind, ira, isis, islamist, johnson, johnson's, jones, labour, labour's, labours, ld, ldem, leanne, lib, libdem, libdems, liberal, manifesto, manifestos, may's, mcdonnell, miliband, miliband's, milliband, minister, mogg, mps, nuttall, obama, oth, palin, paxman, plaid, pm's, policies, policy, poll, polling, polls, portillo, putin, queen's, ref, referendum, republican, republicans, rudd, russia, salmond, sanders, scrapping, sinn, snp, snp's, socialist, soros, sturgeon, sturgeon's, terror, terrorism, terrorist, theresa, thornberry, tns, tories, tories', tory, trident, trump, ukip, ukip's, unionist, vaughan, vladimir, vote, voter, voters, votes, voting, yougov}

\subsubsection{Ambiguous Words}
\seqsplit{
free, \#amazonbasket, \#beastfromtheeast, \#bournemouth, \#colchester, \#eurovision2015, \#finland, \#firearms, \#foxhunting, \#freebies, \#glastonbury, \#grenfell, \#grenfelltower, \#leeds, \#londonbridge, \#marr, \#onelovemanchester, \#spanishgp, \#thursdaythoughts, \#trident, \#uk, \#york, \#youtube, @daraobriain, @emmakennedy, @jamin2g, @janeygodley, @kindleuk, @marcuschown, @paulbernaluk, @pinknews, @rustyrockets, @sophiareed1, @swanseacity1, 10m, 1bn, 2015, 34, 8bn, 8th, allowance, amber, ars, attack, austerity, baptist, behave, benz, branch, burrows, buts, chaos, clarkson, conference, cons, costed, costings, cuts, davey, david, deal, debate, deceased, defend, dementia, diane, diaz, dodging, donald, dunk, ed, emily, esther, exc, factions, fairer, fees, firearms, firefighters, foxes, freddy, general, glastonbury, grenfell, gy, hamlets, hs2, hsbc, ht, htt, hung, inheritance, intention, jeremy, june, kensington, kyle, lab, lambert, landslide, leader, leaders', leaflet, leafleting, leaflets, lorde, marginal, matched, may, mcintyre, meeting, meetup, members, membership, methodist, middlesbrough, middleton, muppets, murphy, natalie, nicola, nigel, observer, participate, party, pensioners, pledge, pledges, plymouth, posters, progressive, prop, reception, register, registered, rehearsing, results, rodgers, rt, ruth, scarpping, scrap, seat, seats, sentencing, sheeran, slater, sos, source, stable, sub, supporting, surge, tactical, tactically, th, theo, tim, transformative, tuition, uc, uk, violence, watson, weak, weir, wheat, youtuber}

\subsection{News case study URL patterns}

\begin{table}[!htb]
\begin{center}
\begin{tabular}{lll}
\toprule
     Source & Type & URL pattern \\ \midrule
Guardian & political & theguardian.com/politics/2018/ \\
& sport & theguardian.com/sport/2018/\\[1em]
BBC & political & bbc.co.uk/news/uk-politics- \\
 & sport & bbc.co.uk/sport/ \\[1em]
Telegraph & political & telegraph.co.uk/politics/2018/ \\
& sport & telegraph.co.uk/football/2018/ \\
& & telegraph.co.uk/cycling/2018/ \\
& & telegraph.co.uk/cricket/2018/ \\
& & telegraph.co.uk/rugby-union/2018/ \\
\bottomrule
\end{tabular}
\caption{URL patterns used to select \textit{sport} and \textit{political} news sharing on Twitter from \textit{The Guardian}, the \textit{BBC}, and \textit{The Telegraph} for the news sharing case study.}
\label{tbl:url-patterns}
\end{center}
\end{table}

\subsection{Full topic models}
We are providing non-political and political topic models in this section.

\subsubsection{Non-Political Topics}

The topics we name in the paper are: 
38 (Entertainment), 
46 (Technology/EU),
48 (Gardening),
58 (MUFC/Premier League), 
75 (London/Beer), 
77 (Follow/Unfollow), 
91 (Africa/Corruption), 
95 (Sugar Tax), 
98 (Premier League), 
114 (Soc. Dem. Values),
148 (Pollution).

The explicitly political topics we identified in the non-political model are 27 (Israel--Palestine), 37 (Scottish independence), 54 (UK politics), 62 (EU politics), 69 (the Green Party), 84 (Brexit), 86 (the US), 106 (Westminster).

\begin{longtable}[!htb]{ p{.05\textwidth}  p{.95\textwidth} } 
            \toprule
         \textbf{Topic} & \textbf{Features}
         \\ \midrule
            1 & risk, daili, de, hr, late, leadership, insur, manag, thetim, innov, aug\\\addlinespace
            2 & youtub, youtub\_video, video, video\_youtub, islam, lyric, hd, feat, music\_video, trailer, video\_make, parodi, muslim\\\addlinespace
            3 & anxieti, ur, mental, mental\_health, depress, mental\_ill, ill, stigma, base, disord, anxious, therapist, mental\_health\_issu, insta, xx\\\addlinespace
            4 & man\_love, steel, andrew, everton, liam, lfc, liverpool, gallagh, jim, lg, oasi, arena, easter\_egg, klopp, giveaway\\\addlinespace
            5 & comic, health, daili, technolog, mentalhealth, healthcar, late, tech, busi, dc, medicin, mm, book\\\addlinespace
            6 & fm, mile, en, gb, step, travel, insid, leadership, eu, construct, stem, sustain\\\addlinespace
            7 & gay, today\_find, pride, tho, netflixuk, porn, coffe, sean, ah, memori, twitter\_account, dublin, domino, work\_today, ugh\\\addlinespace
            8 & ink, illustr, draw, japanes, lesbian, omg, lil, irish, queer, ireland, submiss, dublin, plz, freelanc, eurovis\\\addlinespace
            9 & bronz, trophi, ps, earn, silver, gold, destini, broadcast, assassin, giveaway, gran, game\\\addlinespace
            10 & tourism, student, research, academ, univers, publish, phd, colleg, studi, organis, leisur, philosoph, prof, philosophi, sustain\\\addlinespace
            11 & lewi, xx, haha, chris, happi\_birthday, birthday, gari, cheer, liam, bf, spencer, haha\_good, oop, lol\\\addlinespace
            12 & sampl, repli, bargain, thriller, kindl, free, mysteri, buy, author, book, suspens, amazonuk, trilog, romanc\\\addlinespace
            13 & academi, chariti, budget, kent, educ, sector, trust, tax, manufactur, school, sussex, cyber, mat, implic\\\addlinespace
            14 & chanc\_win, giveaway, competit, follow\_chanc, enter, follow\_chanc\_win, chanc, follow\_win, simpli, prize, winner\_announc, follow\_enter, follow\_retweet, bundl, win\\\addlinespace
            15 & playlist, folk, tea, tuesday, thursday, monday, wednesday, august, decemb, novemb, reel, hardi, rhythm, elvi, octob\\\addlinespace
            16 & angel, pari, sam, bastard, danni, barber, pump, shite, lad, jim, barclay, glenn, innit, daft, casual\\\addlinespace
            17 & ebay, volum, nobl, barn, check, dvd, review, ukchang, sign\_petit, petit, sign\_petit\_ukchang, petit\_ukchang, disc, fossil\\\addlinespace
            18 & fb, mk, past\_week, donat, past, awesom, geek, airport, cake, cycl, browni, amazonuk, phase, protein, workout\\\addlinespace
            19 & autom, schedul, script, web, window, marcus, loop, solut, technic, wizard, stop\_work, scrape, boundari, blog, datum\\\addlinespace
            20 & gt\_gt, gt, gt\_gt\_gt, album, bird, song, music, vibe, mega, moth, dope, make\_sick, bloodi\_good, blur, ala\\\addlinespace
            21 & railway, carriag, rail, destini, train, ps, delay, se, marvel, sigh, molli, virginmedia, berri, geek, lbc\\\addlinespace
            22 & lodg, swansea, ken, swan, leicest, anna, squar, mason, ed, instal, surnam, ceremoni, virginmedia, cardiff\\\addlinespace
            23 & canal, trip, boat, sun, great\_day, festiv, weekend, rugbi, lock, cycl, sail, countrysid, great\_weekend, visitor, temp\\\addlinespace
            24 & sw, pool, waterloo, citi, pr, xxxx, leader, anna, andrea, kyli, compassion, abba, samuel, yawn, grayl\\\addlinespace
            25 & sibl, studi, immigr, patient, diseas, evid, effect, outcom, increas, clinic, diabet, mortal, norm, intervent, genet\\\addlinespace
            26 & shaw, race, bet, cricket, bbcsport, bat, ffs, ball, bowl, counti, fort, moor, hardi, prove\_wrong, tbh\\\addlinespace
            27 & antisemit, israel, jew, palestinian, jewish, isra, racist, anti\_semit, semit, gaza, reluct\\\addlinespace
            28 & post\_photo, photo, facebook, post, fisher, hors, race, xxx, parti, indi, sim, happi\_day, rider, adel, xx\\\addlinespace
            29 & subscript, content, exclus, access, fan, renew, month, join, utd, portug, ab, manchest\_unit, ferguson, leed, man\_utd\\\addlinespace
            30 & durham, newcastl, laura, lauren, xo, hun, isl, dubai, cute, nah, luci, mc, lmao, tbh\\\addlinespace
            31 & develop, birmingham, role, derbi, coventri, net, softwar, job, hire, engin, infrastructur, midland, bbc\_news\\\addlinespace
            32 & ur, pl, gal, wanna, uni, liter, babe, xo, bc, omg, ew, soz, makeup, sophi, casualti\\\addlinespace
            33 & detect, youtub, trailer, cinema, video\_youtub, director, color, magazin, lee, ai, maggi, nasa, arch, depict, vice\\\addlinespace
            34 & imaceleb, doctorwho, wale, gbbo, nurs, swansea, flood, die\_age, doctor, bbcone, phillip, minut\_silenc, iain, evacu, sharon\\\addlinespace
            35 & eurovis, sync, brit, fave, song, spice, fab, chart, singl, battl, woo, wk, love\_song, mexico, album\\\addlinespace
            36 & mi, run, race, marathon, runner, nike, crush, pace, jedi, paul, hm, endur, sprint, gym\\\addlinespace
            37 & scottish, scotland, independ, proport, glasgow, scot, averag, edinburgh, 3, wee, ruth, good\_support, tier, uefa, ranger\\\addlinespace
            38 & beth, xx, haha, xxx, deploy, sean, artist, nicki, wire, hahaha, knit, navi, bbc\_radio, willi, lol\\\addlinespace
            39 & tshirt, york, wed, edinburgh, magic, earn, badg, awesom, film, packag, surf, tee, yorkshir, chanc\_win\\\addlinespace
            40 & keith, trek, samuel, christ, faith, god, spirit, israel, august, promis, ministri, profound, divin, amanda, revel\\\addlinespace
            41 & sheffield, yorkshir, south, vine, centr, ago\_today, year\_ago\_today, sarah, year\_ago, shop, hillsborough, world\_big, arthur, xbox, lol\\\addlinespace
            42 & disabl, dwp, disabl\_peopl, benefit, univers\_credit, poverti, welfar, credit, homeless, auster, pip, sanction, luther, duncan\\\addlinespace
            43 & consult, insight, appl, io, iphon, karl, server, tim, read\_tweet, cook, smartphon, broadband, samsung, softwar\\\addlinespace
            44 & afc, arsenal, wenger, chelsea, spur, arsenal\_fan, alexi, sanchez, tottenham, emir, au, bayern, dortmund\\\addlinespace
            45 & legal, law, court, lawyer, lynn, ms, divorc, student, aid, justic, suprem\_court, profess, mainten, academ\\\addlinespace
            46 & android, dj, ee, eu, app, greec, independ, mobil, scotland, ipad, tutori, currenc, rubi, versus, workout\\\addlinespace
            47 & lt, oxford, input, lt\_lt, hey, stream, ill, nowplay, rank, hall, aha, acoust, meter, gordon, ah\_good\\\addlinespace
            48 & bloom, garden, cute, layer, kinda, ship, charact, rich, bless, hug, hope\_feel, roller, ldn, shoutout, gosh\\\addlinespace
            49 & thanksgiv, km, coffe, cat, tourist, holiday, gym, tube, latin, breakfast, sticker, latt, florida, unlock, kitti\\\addlinespace
            50 & connect, typo, lbc, garylinek, coin, sugar, lord, bot, threaten, climat\_chang, fork, lord\_sugar, edl, man\_make, russian\\\addlinespace
            51 & gay, eurovis, lgbt, money\_make, pride, australia, marriag, equal, sydney, homophob, gay\_man, lgbtq, homophobia, abba, queer\\\addlinespace
            52 & bbc\_news, bbc, news, great\_good, portrait, exhibit, irish, ireland, wonder, museum, serena, twitter\_follow, austria, thame, sad\_news\\\addlinespace
            53 & opinion, fact, agre, understand, genuin, sens, fair, forbid, disagre, interest, logic, necessarili, piti, grasp, satir\\\addlinespace
            54 & bruce, anna, remain, parliament, migrant, illeg, franc, countri, politician, puppi, griev, leaver, traitor, des, imparti\\\addlinespace
            55 & fuck, cunt, fuckin, connor, yer, kyle, shite, ye, shit, lmao, cum, shag, fanni, fuck\_hate, wank\\\addlinespace
            56 & writer, kickstart, movi, film, cinema, write, author, doctorwho, book, episod, haul, andr, crowdfund, day\_leav, big\_screen\\\addlinespace
            57 & follow\_follow, blog, blogger, instagram, shoe, outfit, fashion, dress, gorgeous, beauti, follow\_back, skirt, aso, luxuri, wardrob\\\addlinespace
            58 & mufc, manutd, unit, rooney, manchest\_unit, utd, goal, golf, moy, mourinho, bale, trafford, ucl, cristiano, pogba\\\addlinespace
            59 & shade, film, star\_war, hairdress, starwar, movi, hell, war, jedi, damn, shaun, laser, film\_make, titan, awaken\\\addlinespace
            60 & lol, bald, lmao, im, omg, tho, tbh, ur, ppl, wtf, lol\_good, sooo, goin, carniv, sooooo\\\addlinespace
            61 & cat, eye, eden, mill, insect, fuck\_fuck, biscuit, ian, tree, poem, badger, windsor, good\_god, andr, ha\\\addlinespace
            62 & deficit, el, la, los, remind, econom, en, tax, america, del, es, nationalist, merkel, macron, lo\\\addlinespace
            63 & turkish, lisa, bing, miss, mailonlin, harvey, louis, bloke, lbc, custom\_servic, leo, robbi, vip, xx\\\addlinespace
            64 & art, galleri, artist, exhibit, studio, paint, print, workshop, beach, landscap, contemporari, tide, artwork\\\addlinespace
            65 & cricket, broad, bat, ash, stoke, test, england, bowl, root, ball, aussi, ol, over, jonni, fuck\\\addlinespace
            66 & thor, teacher, physic, activ, educ, child, school, teach, student, earli\_year, nurseri, pe, classroom, toddler, egypt\\\addlinespace
            67 & stain, graphic, websit, van, ff, sign, newcastl, vehicl, updat, wall, saturday\_morn, copper, frost, banner, high\_street\\\addlinespace
            68 & exam, collin, uni, revis, essay, err, sister, lectur, graduat, hull, abbey, gcse, time\_aliv, snapchat\\\addlinespace
            69 & green\_parti, climat, green, climat\_chang, ride, bike, peac, nhs, mag, hunt, pollut, heathrow, countrysid, environment, finland\\\addlinespace
            70 & stamp, paul, writer, morri, blue, andr, scotland, fabric, craft, recip, thumb, sew, stripe, xx, tho\\\addlinespace
            71 & jennif, hahaha, danni, haha, lawrenc, em, il, fav, nottingham, stop\_watch, hunger, larri, fuck, im\\\addlinespace
            72 & market, strategi, content, social\_medium, brand, social, digit, medium, trend, tip, optim, linkedin, influenc, algorithm, facebook\\\addlinespace
            73 & mia, xfactor, sing, factor, song, sterl, anthoni, england, fuck\_fuck, gonna, fluffi, big\_brother, week\_day, carrot, blade\\\addlinespace
            74 & dec, furi, tyson, fight, fighter, box, mate, jr, pal, champ, boxer, ko, parker, big\_man, aj\\\addlinespace
            75 & unit\_kingdom, kingdom, ben, unit, beer, pub, london, januari, station, tim, buffet, slice, high\_street, poet, buff\\\addlinespace
            76 & ppl, didnt, im, sandra, ive, doesnt, isnt, cuz, folk, what, your, hmmm, lol, hes\\\addlinespace
            77 & unfollow, automat, peopl\_follow, stat, follow, check, person, ireland, mention, reach, shane, big\_fan, monitor, vinc, find\_peopl\\\addlinespace
            78 & long\_term, ear, lfc, defend, poor, injuri, midfield, el, defens, hes, dale, trent, analys, salli, your\\\addlinespace
            79 & jane, lt, xxx, xo, nugget, ad, robert, hay, laura, lt\_lt, sob, xxxx\\\addlinespace
            80 & blaze, nfl, wrestl, season, coach, defens, chief, matt, game, offens, vardi, boston, houston, panther, ref\\\addlinespace
            81 & maintain, action, figur, base, statement, appar, talk, link, prefer, give, mental\_health, connect, make, time\\\addlinespace
            82 & council, communiti, wellb, local, west, committe, servic, resid, fund, volunt, social\_care, neighbourhood, district, turnout, great\_hear\\\addlinespace
            83 & nathan, josh, hall, eve, exam, jame, gonna, shit, revis, preston, homework, kennedi, quid, il, fuck\\\addlinespace
            84 & deal, trade, remain, democraci, agreement, union, border, european, freedom, good\_pay, david\_davi, european\_union, leaver, negoti, norway\\\addlinespace
            85 & sign\_petit, petit, frack, ukchang, degre, nhs, petit\_ukchang, sign\_petit\_ukchang, sign, gaza, privatis, open\_letter, georg\_osborn\\\addlinespace
            86 & presid, america, syria, cnn, health\_care, american, gun, senat, white\_hous, potus, congress, fox\_news, presid\_unit, presid\_unit\_state, fbi\\\addlinespace
            87 & ted, steve, neighbour, chris, math, podcast, sum, wealth, hey, parliament, ch, sky\_news, gorilla, puzzl, worth\_watch\\\addlinespace
            88 & barrel, rick, turner, swansea, church, tire, brighton, rugbi, mate, morn, julia, tobi, chick, dissert, briton\\\addlinespace
            89 & chef, restaur, menu, manchest, lunch, food, meal, open, dish, roast, good\_food, dine, norman, citi\_centr, pm\_pm\\\addlinespace
            90 & librari, museum, cat, omg, jo, daniel, liter, ur, staff, pay, patron, windsor, charlott, printer, citizenship\\\addlinespace
            91 & ni, nigeria, lawyer, african, corrupt, parcel, bbcradio, lol, journo, arrest, astronaut, eye\_open, hoo, detain, wk\\\addlinespace
            92 & karaok, drink, gin, bake, tire, cake, nottingham, beer, dave, lil, gemma, cider, burnley, lush, buffet\\\addlinespace
            93 & choir, experi, care, young\_peopl, ne, scotland, wee, kenni, glasgow, edinburgh, portray, peopl\_care, jk\_rowl, show\_support, rowl\\\addlinespace
            94 & carter, saint, billi, st, fa, dan, ranger, bbcsport, band, radio, show\_tonight, nra, curl, di, andi\_murray\\\addlinespace
            95 & frog, level, skill, cunt, sugar, committe, divin, nhs, suspect, holland, jeremi\_hunt, coffin, rag, blog, fascist\\\addlinespace
            96 & theapprentic, ps, nintendo, xbox, switch, mario, episod, game, star\_war, trailer, ea, batman, starwar, consol\\\addlinespace
            97 & raf, york, scotland, poppi, veteran, fraser, soldier, jim, store, boss, primark, lili, today\_rememb, case\_miss, bbcradio\\\addlinespace
            98 & west\_ham, spur, ham, salah, everton, kane, chelsea, klopp, pogba, lukaku, cl, mane, hazard, fuck, midfield\\\addlinespace
            99 & blackpool, wimbledon, itv, liverpool, strict, fabul, tenni, eurovis, ff, manutd, casualti, handsom, watson, saturday\_night, im\\\addlinespace
            100 & vegan, hunt, anim, lash, rescu, wildlif, fox, rickygervai, protect, meat, rhino, endang, extinct, eleph\\\addlinespace
            101 & dart, mark, piersmorgan, lbc, pint, proud, midland, yep, kthopkin, twat, wetherspoon, hood, lap, chuck, rickygervai\\\addlinespace
            102 & leed, veggi, daughter, gbbo, rat, mum, charlott, sew, kim, advert, elton, usernam, crappi, amber, fuck\\\addlinespace
            103 & rio, euro, olymp, gold, medal, doctorwho, bridg, wale, silver, itv, andi\_murray, world\_record, showcas, pursuit, bbcsport\\\addlinespace
            104 & print, font, leed, design, poster, gig, mount, beer, brew, peter, ale, pale, punk, fest\\\addlinespace
            105 & album, gaga, omg, song, ariana, kate, meghan, singl, tour, icon, good\_song, rihanna, banger, beyonc, nicki\\\addlinespace
            106 & economi, rural, britain, unemploy, budget, growth, chancellor, union, bn, westminst, commonwealth, margaret\_thatcher, good\_futur, gdp, booth\\\addlinespace
            107 & bowi, gig, david\_bowi, album, david, band, music, ticket, tour, vinyl, great\_weekend, tix, acoust, happi\_friday, pre\_order\\\addlinespace
            108 & mate, lincoln, lad, haha, newcastl, sunderland, hahaha, aye, dean, cunt, mate\_good, morri, hahahahaha, ashley, fuck\\\addlinespace
            109 & muslim, islam, migrant, rape, tommi, ukrain, religion, gang, tommi\_robinson, immigr, mosqu, free\_speech, hate\_crime, merkel, groom\\\addlinespace
            110 & celtic, ranger, rodger, wit, hes, scottish, bn, robert, brendan, mcgregor, iv, lennon, gerrard, daft, midfield\\\addlinespace
            111 & pin, fanci, newcastl, iphon, favorit, fab, nowplay, nois, bike, wear, christma\_dinner, cube, lamp, rickygervai, bicycl\\\addlinespace
            112 & retail, water, ceo, energi, custom, market, gas, sector, regul, innov, supplier, icymi, storag, resili\\\addlinespace
            113 & yoga, theatr, delici, fabul, cast, workout, drag, mac, audit, dinner, repost, luv, nightclub, popcorn, rehears\\\addlinespace
            114 & foodbank, poverti, nhs, wage, worker, crisi, fund, wick, solidar, food\_bank, grayl, live\_wage, privatis, hunger, comrad\\\addlinespace
            115 & irl, ass, pokemon, damn, movi, dude, meme, bc, charact, shit, wizard, behold, vampir, worm, demon\\\addlinespace
            116 & franci, scout, shift, tattoo, beer, german, tbh, world\_cup, worldcup, incid, navi, ballot, houston, troop, sadiq\\\addlinespace
            117 & black, smh, nah, bro, ass, mad, girl, lmao, drake, tryna, pierc, tl, black\_man, black\_peopl, hoe\\\addlinespace
            118 & cruis, safeti, driver, bbc\_news, fire, leicest, car, vegetarian, member, union, exposur, fleet, trail, england\_wale, fog\\\addlinespace
            119 & world\_cup, goal, cup, fulham, footbal, leagu, worldcup, premier\_leagu, sterl, england, huddersfield, pep, southgat, manchest\_citi, colombia\\\addlinespace
            120 & franchis, agent, properti, episod, bro, soundcloud, ms, review, lol, itun, buyer, estat, mortgag, seller, xx\\\addlinespace
            121 & lfc, liverpool, belfast, klopp, anfield, suarez, xx, hillsborough, gerrard, lui, ski, salah\\\addlinespace
            122 & ironi, racist, rail, remain, ban, odd, invent, lefti, sane, tho, appropri, dodgi, mph, impli, nearbi\\\addlinespace
            123 & panel, energi, solar, hill, climat, local, council, sustain, green, winner, borough, effici, org, st\_centuri, coal\\\addlinespace
            124 & outdoor, mountain, spotifi, lake, peak, trail, summit, district, honey, walk, moss, sculptur, vintag, nowplay, yorkshir\\\addlinespace
            125 & loveisland, georgia, alex, megan, laura, wes, love\_island, liverpool, adam, beyonc, madonna, kyli, loyal, xx\\\addlinespace
            126 & church, fr, pray, mass, bishop, priest, cathol, mari, feast, worship, ministri, christ\\\addlinespace
            127 & villa, aston, cricket, dr, premier\_leagu, xi, premier, england, score, mini, sydney, russia, off, good\_lad, hugh\\\addlinespace
            128 & xxx, fab, xx, xxxx, wonder, love\_show, gorgeous, smile, daughter, ador, make\_smile, daisi, mt, hope\_feel, bravo\\\addlinespace
            129 & photographi, cornwal, photograph, bird, sunset, essex, natur, wolf, photo, van, je, mere, sunris, bonfir, wildlif\\\addlinespace
            130 & carbon, thread, cox, bark, artist, andrea, tran, cc, jersey, energi, orient, fossil, solar, mt, word\_day\\\addlinespace
            131 & calm, london, friday, airport, languag, uniqu, googl, ad, gatwick, german, camden, fur, den, paddington, croatia\\\addlinespace
            132 & welsh, cardiff, wale, languag, member, independ, william, group, english, facebook, renam, mud, royal\_famili, christma\_card, dump\\\addlinespace
            133 & recruit, warehous, graduat, virginmedia, email, glasgow, author, bobbi, web, slight, rupert, ten\_minut, murdoch, amanda, email\_address\\\addlinespace
            134 & mrjamesob, gun, gender, feminist, dumb, argument, dude, disagre, ideolog, nazi, alt, rifl, notch, ration, virtu\\\addlinespace
            135 & tom, jen, tit, nowplay, pint, doo, wrestl, fool, telli, tbh, minc\_pie, minc, preston, fuck\\\addlinespace
            136 & christian, bibl, jesus, christ, ye, church, sin, pray, god, faith, satan, worship, vers, atheist, flesh\\\addlinespace
            137 & bristol, cook, bang, tape, record, arm, releas, bear, hideous, field, bleach, trace, gag, twelv, cd\\\addlinespace
            138 & googl, search, index, mobil, site, rank, content, tool, updat, link, algorithm, thx, optim, crawl, fetch\\\addlinespace
            139 & clarkson, cycl, favorit, eastend, driver, pride, bike, alan, gun, cyclist, waitros, potus, adida, paedophil, bicycl\\\addlinespace
            140 & bed, jodi, sleep, boyfriend, tire, flavour, eat, wash, nap, tesco, bedtim, payday, bra, lose\_weight, crave\\\addlinespace
            141 & io, appl, raspberri, app, microsoft, iphon, window, ipad, code, mac, chrome, beta, browser, os, butler\\\addlinespace
            142 & airway, jet, nowplay, flight, rout, airport, leed, pepper, ryanair, airlin, paradis, ba, heathrow\\\addlinespace
            143 & norwich, neil, wes, gut, ticket, bet, championship, superb, refere, season, game\_today, season\_ticket, alloc, applaus, win\_ticket\\\addlinespace
            144 & ha\_ha, ha, wright, nurs, placement, ami, jo, disabl, luke, hug, brace, ar, jo\_cox, orlando, lol\\\addlinespace
            145 & korean, riot, na, player, team, split, tournament, korea, region, champ, passiv, reddit, tier, leagu, game\\\addlinespace
            146 & gender, tran, feminist, sex, rape, woman, sexual, male, femal, femin, gum, misogyni, sexism, prostitut, sexual\_assault\\\addlinespace
            147 & greg, cfc, chelsea, cont, sadiqkhan, iran, racist, mourinho, hazard, staff, diego, fifti, grenfelltow, fuck, jose\\\addlinespace
            148 & plastic, contain, ocean, pollut, dr, usa, sea, recycl, dwp, path, curios, litter, tonn, live\_uk, crook\\\addlinespace
            149 & southampton, player, saint, squad, club, fc, season, midfield, loan, leagu, crook, start\_season, make\_chang, scorer, delight\_announc\\\addlinespace
            150 & blake, chees, babe, holli, dale, aha, haha, joe, ahh, mac, fml, aso, dissert, deffo, bruis\\\addlinespace
        \bottomrule
        \caption{Words belongs to non-political topics. Unique words are selected by combining top 10 words of FREX, LIFT and log scores in given order.}
    \label{tbl:topic_words_nonpolitical}
\end{longtable}

\subsubsection{Political Topic Model}

This is the model created with the political text dataset.

\begin{longtable}{ p{.05\textwidth}  p{.95\textwidth} } 
    \toprule
         \textbf{Topic} & \textbf{Features}
         \\ \midrule
            1 & blue, chicken, stori, ukchang, thegreenparti, vote\_green, food, stuff, breakfast, philip, fri, lover, elect\_manifesto, clair, green\\\addlinespace
            2 & life, world, fulfil, promis, breakdown, announc, labour, admir, check, gap, precis\\\addlinespace
            3 & mate, teen, fiction, stori, wolf, truth, true\_stori, beach, heart, dirti, stori\_good, top\_stori, love\_stori\\\addlinespace
            4 & tomorrow, tonight\_vote, tonight, vote\_tomorrow, forget, sum, asham, home, centr, releas, vote\_govern, liam, mp\_back, week\_ago, vote\\\addlinespace
            5 & jeremycorbyn, marri, jeremycorbyn\_theresa, peopl, jeremycorbyn\_labour, stand, make, peopl\_vote, vote\_ge, sad, theresa\_jeremycorbyn, behav, neighbour, vote\_jeremycorbyn, februari\\\addlinespace
            6 & green, parti, polit\_parti\_side, green\_parti, today, brand, campaign, labour\_back, labour\_polit, labour\_polit\_parti, parti\_ge, conserv\\\addlinespace
            7 & prime, prime\_minist, minist, categori, presid, live, chief, birthday, call, offic, sincer, minist\_theresa, life\\\addlinespace
            8 & soar, vote\_retweet, retweet, sampl, retweet\_vote, maguir, sampl\_size, kevin\_maguir, size, kevin, dailymailuk, tomorrow\_vote, poll, faisalislam\\\addlinespace
            9 & bbcqt, labour\_ge, hrtbps, audienc, bbcquestiontim, owenjon, dimblebi, carolineluca, soubri, anna\_soubri, interrupt, hmm, audienc\_member, white\_man, vote\_record\\\addlinespace
            10 & agre, matter, true, wrong, politician, vote, vote\_vote, peopl\_vote, chang, vote\_matter, vote\_parti, parti\_vote, chang\_vote\\\addlinespace
            11 & ref, vote\_support, cheer, lee, wow, morn, appoint, mr, jack, laugh, coach, merri\_christma, moan, asap, milliband\\\addlinespace
            12 & chanc\_win, competit, enter, select, follow, winner, chanc, product, prize, copi, tag, entri, fanci, win\\\addlinespace
            13 & cameron, david\_cameron, david, miliband, surgeri, ed, ed\_miliband, telegraph, osborn, pay, reshuffl, georg\_osborn, labour\_leader, pmqs\\\addlinespace
            14 & gb, york, endur, sajidjavid, javid, ruthdavidsonmsp, squar, arrang, remoan, good\_luck, oop, curious, sajid, corbyn\\\addlinespace
            15 & instagram, snapchat, girl, storylin, artist, gonna, music, fuck, favorit, song, cute, good\_stori, album, horror\_stori, true\_stori\\\addlinespace
            16 & jean, william, plaid, lib\_dem, lib, leann, dem, cardiff, bbcdebat, ld, erm, bad\_idea, leannewood, mare, tori\_major\\\addlinespace
            17 & rhetor, energi, water, sector, industri, connect, blog, invest, berlin, confer, ceo, sustain, retail, cc, leadsom\\\addlinespace
            18 & read, small, outrag, act, make\_big, differ, big, luck, govt, power, wash, mouth, yep, door, love\\\addlinespace
            19 & bbc, bbc\_news, make\_happen, news, davi, bbcnew, david\_davi, news\_brexit, blair, fish, news\_tori, river, banner, botch, brexit\\\addlinespace
            20 & revis, exam, histori, ur, uni, lol, god, haha, obama, annoy, horror\_stori, histori\_book, toy\_stori, presid\_obama, snapchat\\\addlinespace
            21 & clegg, nick\_clegg, nick, send\_messag, leadersdeb, mind, divis, tuition\_fee, end, tuition, xenophobia, sin, childish, generous, goodby\\\addlinespace
            22 & hollywood, front\_page, page, newspap, front, soldier, beer, host, soviet, journal, hillsborough, genocid, murray, flee, instal\\\addlinespace
            23 & septemb, vote, labour, check, card, life, announc, world, wild, stori, dog\\\addlinespace
            24 & support\_peopl, owner, shop, pride, mine, wolf, oop, agent, wall, remain\_win, govern\_support, peopl\_peopl, merri\_christma, support, wto\\\addlinespace
            25 & carolin\_luca, liberti, guidofawk, telegraphnew, brighton, compromis, fish, afneil, carolin, dpjhodg, clap, luca, retail, cricket, wilson\\\addlinespace
            26 & peoplesvot, peoplesvot\_uk, ordinari\_peopl, martin, insult, edinburgh, deliv, outcom, attempt, approach, bell, call\_peopl, bbc\_live, bad\_brexit, uk\\\addlinespace
            27 & conserv, vote\_conserv, parti\_side, conserv\_parti, conserv\_win, parti\_work, parti, conserv\_mp, conserv\_govern, today, good\_good, conserv\_vote, campaign, polit\\\addlinespace
            28 & sign\_petit, ukchang, petit, sign\_petit\_ukchang, sign, petit\_ukchang, theresa\_mp, anim, hunt, farm, trophi, xx, petit\_call, call\_govern\\\addlinespace
            29 & bold, elect\_quiz, quiz, uk\_elect, elect\_quiz\_result, quiz\_result, wsyvf, uk\_elect\_quiz, result\_labour, result, exhibit\\\addlinespace
            30 & rend, properti, agenc, photograph, landlord, insur, provid, embrac, agent, inform, cloud, tenant, aspect, packag, vat\\\addlinespace
            31 & leadersdeb, russel\_brand, libdem, bbcdebat, barackobama, farron, tim\_farron, ge, candid, liber\_democrat, natali\_bennett, hust, charl, natali, elect\_debat\\\addlinespace
            32 & technolog, comic, health, daili, late, health\_care, busi, tech, con, conserv\_win, ep, top\_stori, great\_stori, stori\\\addlinespace
            33 & kent, scienc, plastic, environ, natur, climat, theatr, scheme, nato, beauti, recycl, museum, marin, robot, exhibit\\\addlinespace
            34 & life, check, promis, world, stori, life\_stori, announc, vote, true, match, wed\\\addlinespace
            35 & gibraltar, collabor, spanish, wto, escap, spain, freedom, de, sovereignti, evil, 4, convent, dictatorship, anti\_democrat, oppress\\\addlinespace
            36 & lib\_dem, lib, dem, vote\_futur, vote\_lib\_dem, vote\_lib, lol, tim, amber\_rudd, rudd, snake, yup, gear, comic, sigh\\\addlinespace
            37 & cancer, healthi, test, research, mum, amaz, share, inspir, share\_stori, birmingham, top\_stori, fab, love\_stori, charli, grammar\_school\\\addlinespace
            38 & hiv, experi, student, care, sexual, woman, welsh, lgbt, young, inspir, share\_stori, launch\_campaign, sister, peopl\_live, harass\\\addlinespace
            39 & salmond, indyref, telegraph, shot, scot, currenc, scottish\_independ, scottish, scotland, independ, milliband, scotland\_vote, sterl, undecid, snp\\\addlinespace
            40 & battlefornumb, nhsmillion, leadersdeb, doctor, nhs, dr, jeremi\_hunt, ge\_labour, paxman, privatis, exempt, privatis\_nhs, uk\_polit, ukchang, dwp\\\addlinespace
            41 & unbeliev, wit, ref, thick, pathet, folk, disgrac, newsnight, pr, daili\_mail, fiasco, carpet, slaveri, year\_year, bitter\\\addlinespace
            42 & termin, cornwal, googl, comput, datum, facebook, art, search, site, websit, storytel, web, museum, traffic, exhibit\\\addlinespace
            43 & polit, parti, polit\_parti, campaign, today, judg, debat, mp, reform, system, parti\_leader, british\_polit, conserv\\\addlinespace
            44 & manchest, wisdom, uklabour, abbott, wood, dian\_abbott, dian, stephen, euref, resid, grammar\_school, grammar, blast, champagn, steven\\\addlinespace
            45 & reward, swindon, pool, leadsom, andrea, rachael, rachael\_swindon, wale, voteleav, vote\_leav, compassion, badg, dpjhodg, britainelect\\\addlinespace
            46 & isi, saudi, yemen, obama, plea, syria, clinton, refuge, donald\_trump, arabia, hunger, humanitarian, barack\_obama, presid\_obama, barack\\\addlinespace
            47 & vote\_lose, irish, im, gonna, magic, abort, tree, money\_tree, magic\_money\_tree, magic\_money, dublin, wanna, black\_woman, lend, cathol\\\addlinespace
            48 & gt, gt\_gt, gay, marriag, miss, facebook, news, blog, sex, datum, privaci, berlin, nake, sex\_marriag, bake\\\addlinespace
            49 & singl\_market, hard\_brexit, procedur, davi, brexit\_brexit, market, brexit, singl, citizen, negoti, brexit\_impact, ken\_clark, ep, eea, soft\_brexit\\\addlinespace
            50 & disabl, welfar, laboureoin, welfar\_cut, poverti, jon, cut, mirrorpolit, swindon, tori, tax\_credit, corbyn\_leader, bedroom\_tax, dailymirror, harryslaststand\\\addlinespace
            51 & innov, guardian, ian, drink, creativ, advertis, art, ad, food, de, storytel, gif, ted, magazin\\\addlinespace
            52 & peoplesvot, stopbrexit, proport, brexit\_deal, fbpe, represent, deal\_brexit, squeez, democraci, stop\_brexit, vote\_system, make\_voic\_hear, miser, peoplesvotemarch, brexit\\\addlinespace
            53 & ree\_mogg, mogg, ree, bori\_johnson, bori, jacob, jacob\_ree\_mogg, jacob\_ree, johnson, dear\_theresa, shower, tobi, call\_jeremi, michael\_gove, cartoon\\\addlinespace
            54 & cambridg, council, stark, councillor, fbpe, resid, citi, amend, motion, local\_elect, scrutini, final\_deal, localelect, vote\_local, district\\\addlinespace
            55 & ft, victoria, iandunt, folk, music, tuesday, monday, wednesday, podcast, bill, februari, final\_brexit\_deal, final\_brexit, hillaryclinton, guardian\\\addlinespace
            56 & launch, campaign, today, commit, polit, parti, confer, pledg, green, futur, parti\_confer\\\addlinespace
            57 & councillor, local, joy, council, tim, canvass, candid, resid, cpc, ward, borough, district, afternoon, session, local\_govern\\\addlinespace
            58 & welsh, ceremoni, wale, cymru, cardiff, plaid, dawn, deputi, plaid\_cymru, councillor, deputi\_leader, local\_govern\\\addlinespace
            59 & elector\_commiss, custom\_union, brussel, custom, commiss, 2\_referendum, chequer, elector, trade, union, darren, remain\_campaign, grime, reckon, regulatori\\\addlinespace
            60 & fb, regist\_vote, regist, young\_peopl, ireland, northern\_ireland, northern, girl, young, obligatori, peopl\_regist, make\_voic\_hear, mini, make\_voic, young\_peopl\_vote\\\addlinespace
            61 & comprehens, smear, tori\_mp, jeremi\_corbyn, video, msm, accus, expos, jeremi, homeless, bomb\_syria, blairit, nerv, attack\_corbyn, huffpostukpol\\\addlinespace
            62 & deal\_brexit, ree, mogg, ree\_mogg, deal, jacob, erg, jacob\_ree, jacob\_ree\_mogg, brexit\_deal, raab, domin\_raab, irish\_border, white\_paper, countri\_brexit\\\addlinespace
            63 & skynew, ukip\_vote, skynewsbreak, gold, airport, hour, leed, maguir, idiot, abbott, olymp, flight, maker, yorkshir, kevin\_maguir\\\addlinespace
            64 & vote\_labour, labour\_parti, labour\_win, labour, vote\_labour\_vote, labour\_govern, votelabour, dup, tori\_govern, labour\_vote, theresa\_lose, lose\_major, labour\_gain, labour\_elect, labour\_labour\\\addlinespace
            65 & realdonaldtrump, piersmorgan, potus, crook, presid, trump, nigel\_farag, nigel, presid\_trump, fake, anti\_trump, haha, cnn, william, fbi\\\addlinespace
            66 & today\_good, id, bbclaurak, king, enter, broadcast, code, user, text, garden, santa, privaci, marvel, februari, communic\\\addlinespace
            67 & battl, christma, brit, rais, round, total, chart, favourit, girl, video, relief, freez, itv, bag, storm\\\addlinespace
            68 & auster, nurs, nhs, brexit\_govern, homeless, poverti, jeremi\_hunt, windrush, disabl, grenfel, live\_poverti, underfund, health\_secretari, hostil\_environ, univers\_credit\\\addlinespace
            69 & safeti, strong, pmqs, stabl, strong\_stabl, good\_brexit, stabl\_leadership, strong\_stabl\_leadership, bbcqt, economi, plan\_brexit, coalit\_chao, download, queensspeech, strengthen\\\addlinespace
            70 & voteleav, vote\_leav, euref, leav\_eu, tourist, vote\_leav\_eu, leav, june, eu\_referendum, vote\_brexit, gareth, knight, vote\_stay, eureferendum, opt\\\addlinespace
            71 & snp, nicola\_sturgeon, nicola, sturgeon, shortlist, salmond, scottish, labour\_snp, scotland, alex, vote\_snp, independ\_referendum, ge, ruthdavidsonmsp\\\addlinespace
            72 & antisemit, jew, jewish, israel, aaronbastani, pander, anti\_semit, holocaust, semit, smear, conspiraci\_theori, conspiraci, uklabour\_jeremycorbyn, support\_corbyn, uklabour\\\addlinespace
            73 & fuck, shit, tempt, utter, bastard, piss, genuin, ffs, kind, bbcnew, give\_shit, dude, horror\_stori, wanker, bbc\_live\\\addlinespace
            74 & labour\_leader, complac, labour\_parti, corbyn\_labour, manifesto, cooper, yvettecoopermp, corbyn, labour\_mp, lib\_dem, yvett\_cooper, yvett, trust\_tori, clap, ferri\\\addlinespace
            75 & side\_labour, drink, vote\_vote, mum, nurs, strong\_stabl, band, paxman, rob, stabl, toy\_stori, wheat, nurs\_pay, terrorist\_sympathis, toy\\\addlinespace
            76 & mail, royal, worker, strike, union, offic, privatis, post, postal, pmqs, deliveri, junior, probe, murdoch, sign\_petit\\\addlinespace
            77 & parti\_countri, ld, snp, scottish, miliband, faisalislam, tori, swing, tori\_labour, labour\_tori, stanley, tori\_lib, tori\_seat, gordon\_brown, boundari\\\addlinespace
            78 & australia, senat, unexpect, australian, liber, scott, climat, feder, alan, motion, august, fox\_news, craig, coal, trump\\\addlinespace
            79 & vest, cunt, bbcnew, polic, bbcpolit, sun, aid, berni, syria, vote\_ukip, greet, keith, oversea, spit, killer\\\addlinespace
            80 & trump, gop, republican, presid\_trump, senat, putin, presid, fact\_check, white\_hous, cohen, mueller, fox\_news, trump\_tweet, congress, trump\_trump\\\addlinespace
            81 & owenjon, uklabour, tide, iandunt, resign, starmer, keir\_starmer, keir, borisjohnson, bbcradio, numb\_gov, downingstreet, bbcnewsnight, vote\_uklabour, uklabour\_jeremycorbyn\\\addlinespace
            82 & youtub, video, guard, lbc, mrjamesob, reform, stephen, elector, system, sander, berni\_sander, peopl\_vote\_tori, brien, vote\_system, grime\\\addlinespace
            83 & disabl, carolineluca, labour\_seat, palestinian, disabl\_peopl, auster, pr, israel, poverti, mrjamesob, lloyd, brighton, dwp, apartheid, univers\_credit\\\addlinespace
            84 & islam, dump, christian, isi, muslim, potus, god, robinson, mr, liber, islamophobia, tommi\_robinson, feminist, anti\_trump, leftist\\\addlinespace
            85 & rowl, jk, jk\_rowl, seeker, 2\_referendum, bath, ben, leaver, voic, norway, duck, echo, chap, coordin, snp\\\addlinespace
            86 & make\_bad, dan, argument, oppos, iandunt, remain, yep, kind, fli, assum, notion, sceptic, se, narrow, reassur\\\addlinespace
            87 & sunday, articl, view, today, stand, peopl\_vote, make, peopl, interview, tonight, love, vote\\\addlinespace
            88 & gun, bc, gari, liber, canada, ford, canadian, pc, gop, wing, tend, ride, ownership, govt\\\addlinespace
            89 & barri, leaver, sean, jon, mac, david, precious, jolyonmaugham, eu, rachael\_swindon, violat, permit, peoplesvot\\\addlinespace
            90 & anna, theresamay, good\_theresa, theresa, battlefornumb, brilliant, leadersdeb, dian\_abbott, borisjohnson, dian, stewart, time\_theresa, patel, theresa\_theresa, leann\\\addlinespace
            91 & ukip, vote\_ukip, nigel\_farag, nigel, lt, farag, time\_tori, chairman, establish, style, ukip\_candid, ukip\_leader, steven, ukip\_support, januari\\\addlinespace
            92 & histori\_make, ww, movi, railway, cymru, dad, bird, volunt, histori, beach, cafe, mill, museum, compens, leftist\\\addlinespace
            93 & governor, school, budget, chariti, fraud, click, math, educ, teacher, detail, autumn, tight, audit\\\addlinespace
            94 & semit, anti\_semit, corbyn, owen, antisemit, momentum, labour\_parti, jeremi\_corbyn, jeremi, labour\_mp, parti\_mp, nec, corbynit, newcastl, denier\\\addlinespace
            95 & chequer, telegraph, telegraphnew, backstop, rachel, borisjohnson, cabinet, guidofawk, custom\_union, bori, commonwealth, brexit\_plan, mailonlin, loyal, conserv\_manifesto\\\addlinespace
            96 & if, starmer, tran, keir, keir\_starmer, lgbt, peston, reach, amend, south, thesun, ge\_labour, untru, inherit, ge\\\addlinespace
            97 & eurovis, ps, song, leagu, team, duti, player, riot, game, semi, format, stream, alcohol, org, tan\\\addlinespace
            98 & leagu, player, season, club, victori, arsenal, footbal, goal, ref, game, eng, premier\_leagu, premier, full\_stori, fa\\\addlinespace
            99 & real\_world, storylin, advert, brexitchao, dog, tommi, danc, male, son, ring, brag, tan, chocol, presid\_unit, presid\_unit\_state\\\addlinespace
            100 & muslim, islam, terrorist, extent, terror, ira, rape, flag, terror\_attack, irish, mosqu, victorial, teresa, white\_peopl, hate\_crime\\\addlinespace
            101 & christian, prayer, church, syria, syrian, jesus, isi, town, christ, pray, grace, middl\_east, israel\\\addlinespace
            102 & mrjamesob, lbc, femi, cunt, realdonaldtrump, ouch, bbcnew, jacob\_ree, jacob\_ree\_mogg, jacob, krishgm, downingstreet, adoni, fbpe, marrshow\\\addlinespace
            103 & black, founder, black\_woman, lol, african, black\_peopl, inspir, emot, cultur, woman, michell\_obama, creativ, privaci, michell, stori\\\addlinespace
            104 & ge\_ge, palestinian, israel, isra, gaza, hama, votelabour, jam, palestin, jc, occupi, mandela, jr, jc\_pm, voteleav\\\addlinespace
            105 & studi, imperi, effect, immigr, neoliber, isi, econom, thread, western, intellig, nativ, presid\_unit, presid\_unit\_state, explicit, civilian\\\addlinespace
            106 & nicolasturgeon, snp, thesnp, scotland, scottish, glasgow, indyref, independ, nicola, scot, vote\_snp, indi, independ\_referendum, ruthdavidsonmsp, tori\\\addlinespace
            107 & jc, jc\_pm, rachael\_swindon, rachael, news\_theresa, swindon, jeremi\_corbyn, angelarayn, theresa, jeremi, toryfib, el\_jc, harryslaststand, jamesmelvill, vote\_jeremycorbyn\\\addlinespace
            108 & yougov, wale, lab, conserv\_labour, lead, britainelect, gain, unemploy, cardiff, ld, local\_council, labour\_lib\_dem, labour\_lib, tori\_lead, theresa\_conserv\\\addlinespace
            109 & loud, languag, english, shout, play, continu, show, polici, store, prefer, sing, judg\\\addlinespace
            110 & hillari, clinton, hillaryclinton, hillari\_clinton, forget\_vote, electionnight, trump\_win, donald, presid, donald\_trump, vote\_poll, electionday, ivot, trump\\\addlinespace
            111 & south\_africa, africa, cpc, south, miliband, ed, itvnew, bbcpolit, economi, marrshow, good\_futur, michell, recess, queensspeech, labour\\\addlinespace
            112 & short\_stori, short, book, episod, film, write, writer, seri, charact, librari, ghost, stori\_time, love\_stori, great\_stori, storytel\\\addlinespace
            113 & call\_referendum, libdem, final\_brexit, final\_brexit\_deal, brexit\_deal, amend, final, referendum, liber\_democrat, join, brexit\_impact, stopbrexit, brexit, fbpe, timfarron\\\addlinespace
            114 & hotel, donald\_trump, donald, cymru, wale, reveal, euro, bridg, presid\_trump, trump, moon, jong, kim\_jong, palac, spit\\\addlinespace
            115 & wto, leav\_eu, surrend, vote\_leav, remain, remoan, custom\_union, trade, brussel, vote\_leav\_eu, million\_vote, democrat\_vote, everyday, project\_fear, peopl\_vote\_leav\\\addlinespace
            116 & wto, soubri, guidofawk, bbcdp, anna\_soubri, deal\_brexit, anna, deal, leav\_eu, bercow, brexit\_mp, deselect, constitu\_vote, andrew\_neil, withdraw\_agreement\\\addlinespace
            117 & ge\_match, parti\_ge\_match, green\_parti\_ge, match, votematch\_match\_green, match\_green\_parti, parti\_ge, lgbt, match\_green, gay, sex\_marriag, album\\\addlinespace
            118 & votematch\_match, newsnight, ge, bbcdebat, votelabour, lt, generalelect, electionnight, eurovis, bbcelect, trumpet, newcastl, ge\_labour\\\addlinespace
            119 & votelabour, today\_theresa, forthemani, jeremycorbyn\_labour, laboureoin, jeremycorbyn\_pm, berni, jeremi\_corbyn, rachael, rachael\_swindon, berni\_sander, texa, labour\_gain, theresa\_resign, florida\\\addlinespace
            120 & thegreenparti, bristol, carolineluca, anim, natalieben, vote\_green, climat, green\_parti, environ, frack, green, luca, carolin\\\addlinespace
            121 & libdem, timfarron, lib\_dem, liber\_democrat, lib, dem, liber, citizen, eu\_citizen, democrat, rebuild, vincec, edinburgh, polish, freedom\_movement\\\addlinespace
            122 & bbcr, bbcr\_today, bbcnew, prejudic, fbpe, chunkymark, bbcradio, britain, davi, british\_peopl, tobi, porn, stockpil, peoplesvot, brexit\\\addlinespace
            123 & side, make, polici, campaign, find, feel, peopl, back, work, great, view, futur, love, bite, vote\\\addlinespace
            124 & votematch, joe, match, climat\_chang, wage, climat, scrap, alcohol, militari, student, govern\_cut, billi, cafe, low\_pay, dailymailuk\\\addlinespace
            125 & laboureoin, amberruddhr, aaronbastani, angelarayn, tori, neoliber, socialistvoic, auster, chunkymark, nhs, toryfib, tori\_nhs, barrygardin, corbyn\\\addlinespace
            126 & jeremycorbyn\_uklabour, antisemit, jew, jewish, israel, uklabour, peoplesmomentum, dpjhodg, isra, hama, discriminatori, uklabour\_jeremycorbyn, examin, corbynista, sunni\\\addlinespace
            127 & fabric, posh, mike, rob, paulmasonnew, johnmcdonnellmp, owenjon, ww, agent, dpjhodg, jamesmelvill, portug\\\addlinespace
            128 & miliband, ed, ed\_miliband, uklabour, tax, small\_busi, cchqpress, david\_cameron, unemploy, nhs, tax\_avoid, kitchen, vote\_uklabour, tax\_rise, snp\\\addlinespace
            129 & penalti, world\_cup, cup, russia, goal, footbal, club, leagu, worldcup, chelsea, premier\_leagu, jami, liverpool, cunt, fuck\\\addlinespace
            130 & peoplesvot, cheat, ten\_thousand, investig, crimin, arron, chang\_mind, arron\_bank, stopbrexit, elector, cambridg\_analytica, analytica, elector\_law, break\_law, charlatan\\\addlinespace
            131 & nomin, full, vote\_today, st, vote, join, tonight, stori, link, retweet, cast\_vote, joe, special, campaign, today\\\addlinespace
            132 & opinion, view, stand, peopl, peopl\_vote, disagre, articl, countri, poor, make, black, socialist, excus\\\addlinespace
            133 & manufactur, lbc, norway, make\_hard, hard\_brexit, car, export, skynew, dpjhodg, soft, boil, hoc, soft\_brexit, ultra, ken\_clark\\\addlinespace
            134 & con\_lab, lab\_ldem, ldem, con\_lab\_ldem, vote\_intent, lab, westminst\_vote, westminst\_vote\_intent, ldem\_ukip, con, cut\_tax, grn\\\addlinespace
            135 & squad, fan, player, tho, ref, ridicul, game, score, rich, club, didnt, baffl, fa, pat, loan\\\addlinespace
            136 & heaven, london, mayor, stephen, exit\_poll, bori, queen, highlight, standardnew, khan, theatr, cow, burnham, theresa\_jeremycorbyn, today\_day\\\addlinespace
            137 & twitter\_poll, lbc, timfarron, hillari, fuck, twat, dian, owen, owenjon, owen\_jone, mental\_ill, fav, snowflak, trump\_presid, tit\\\addlinespace
            138 & uk\_govt, petit, sign\_petit, toriesout, sign, frack, climat, eu\_citizen, degre, plastic, petit\_call, bee, open\_letter, coal, call\_govern\\\addlinespace
            139 & bid, reject, fals, trust, leader, true, sound, confid\_vote, mind, care, merri, win\_back, naiv, spokesman, mother\\\addlinespace
            140 & 5, dan, powel, skinner, frank, review, neil, yorkshir, gentl, peoplesmomentum, distanc, bake, bath, portug\\\addlinespace
            141 & proud, labour, hope, labour\_parti, love, tonight, fight, announc, futur, countri, sad, wonder, dedic, vote, speech\\\addlinespace
            142 & ira, ta, mcdonnel, rant, semit, jeremi\_corbyn, murder, owen\_jone, anti\_semit, rail, jezza, emma, emilythornberri, remot, moron\\\addlinespace
            143 & theme, campbellclaret, internet, gender, im, victim, telegraphnew, anim, davidlammi, digit, supermarket, org, tooth, justin, anonym\\\addlinespace
            144 & crap, hackneyabbott, davidlammi, sadiqkhan, lol, bbcbreak, jeremycorbyn\_theresa, uklabour, rubbish, jesus, clueless, prick, um, jeremycorbyn, corbyn\\\addlinespace
            145 & shitti, davidlammi, sugar, lord\_sugar, iran, hahaha, kevin, shambl, maguir, kevin\_maguir, damian, lose\_job, bbcnew\\\addlinespace
            146 & eu, good\_eu, leav\_eu, european, trade, border, europ, migrat, trade\_deal, pro\_eu, britain\_leav, greec, countri\_eu, join\_eu, eu\_eu\\\addlinespace
            147 & aye, pint, cunt, fuck, simon, neil, georgeaylett, vote\_tori, brexit\_brexit, ha, fuck\_tori, telli, billi, ma, tori\\\addlinespace
            148 & putin, russian, resolv, ukrain, russia, lbc, clown, telegraphnew, sanction, muslim, eastern, vladimir\_putin, moscow, mayor\_london, vladimir\\\addlinespace
            149 & standardnew, guardian, milk, realdonaldtrump, gun, bbcnew, telegraph, abort, jeremycorbyn\_pm, skynew, wee, cop, cure, witch\\\addlinespace
            150 & la, gop, republican, senat, de, en, cnn, el, moor, democrat, center, roy, trump, liber\\\addlinespace
        \bottomrule
\caption{Words belongs to political topics. Unique words are selected by combining top 10 words of FREX, LIFT and log scores in given order.}
 \label{tbl:pol_words}
\end{longtable}


\end{document}